\documentclass[10pt,twocolumn,twoside]{IEEEtran}

\ifCLASSINFOpdf

\else
 
\fi

\usepackage{latexsym}
\usepackage{graphicx}
\usepackage{amssymb}
\usepackage{amsmath}
\usepackage{subfigure}
\usepackage{stmaryrd}
\usepackage{cite}

\begin{document}

\title{Quasi-Optimal Network Utility Maximization for Scalable Video Streaming}

\author{\IEEEauthorblockN{Mohammad Sadegh Talebi*,
Ahmad Khonsari, Mohammad H. Hajiesmaili, Sina Jafarpour}
\thanks{M. S. Talebi is with the School of Electrical Engineering, The Royal Institute of Technology (KTH), 100 44, Stockholm, SWEDEN (email: mstms@kth.se).

A. Khonsari and M. H. Hajiesmaili are with Electrical and Computer Engineering Department of University of Tehran, and with the School of Computer Science, Institute for Research
in Fundamental Sciences (IPM), Niavaran Sq., Tajrish Sq., Tehran, IRAN, P. O. Box
19395-5746 (email: ak@ipm.ir, hajiesmaili@ipm.ir).

S. Jafarpour is with Department of Computer Science, Princeton University, 35 Olden Street, Princeton, NJ 08540-5233 (email: sina@cs.princeton.edu).}}

\maketitle

\begin{abstract} 
This paper addresses rate control for transmission of scalable video streams via Network Utility Maximization (NUM) formulation. Due to stringent QoS requirements of video streams and specific characterization of utility experienced by end-users, one has to solve nonconvex and even nonsmooth NUM formulation for such streams, where dual methods often prove incompetent. 
\emph{Convexification} plays an important role in this work as it permits the use of existing dual methods to solve an approximate to the NUM problem iteratively and distributively. Hence, to tackle the nonsmoothness and nonconvexity, we aim at reformulating the NUM problem through approximation and transformation of the ideal discretely adaptive utility function for scalable video streams. The reformulated problem is shown to be a D.C. (Difference of Convex) problem. We leveraged Sequential Convex Programming (SCP) approach to replace the nonconvex D.C. problem by a sequence of convex problems that aim to approximate the original D.C. problem.  
We then solve each convex problem produced by SCP approach  using existing dual methods. This procedure is the essence of two distributed iterative rate control algorithms proposed in this paper, for which one can show the convergence to a locally optimal point of the nonconvex D.C. problem and equivalently to a locally optimal point of an approximate to the original nonconvex problem. Our experimental results show that the proposed rate control algorithms converge with tractable convergence behavior. 
\end{abstract}


\IEEEpeerreviewmaketitle

\newtheorem{myLemma}{Lemma}
\newtheorem{myTheo}{Theorem}
\newtheorem{myDef}{Definition}
\newtheorem{myProp}{Proposition}

\section{Introduction}
\label{sec:intro}
Over the past decade, there has been a rapidly increasing interest in multimedia applications in networking paradigms. Video-based applications such as live-streaming and video conferencing are in possession of the dominant share of such applications. 
Due to necessity of stringent QoS requirements, video transmission proves quite challenging. Throughput variation that often occurs in both wired and wireless networks even exacerbates the problem. While this phenomenon is an inherent property of wireless networks that occurs due to fading and shadowing, wired networks also experience it as a result of network congestion \cite{schaar2005cross,zhang2004end}. 

Video adaptation schemes have been proposed to evince efficient means, not only for rate control to combat against throughput variations, but also to tune video quality to terminal capability and user preference on a per-user basis \cite{ChangVideoAdaptation}. 
As one of the most efficient video adaptation schemes, Scalable Video Coding (SVC) scheme allows for encoding a high-quality video bitstream that
contains one or more valid and decodable subset bitstreams \cite{OhmScalable,SchwarzScalable}. As an extension to H.264/AVC standard, SVC remedies challenges in video transmission through temporal, spatial, and quality (PSNR) scalability of the video stream, resulting in exhibition of several quality classes.

To date rate allocation for video streaming over wired and wireless networks has been studied extensively \cite{Dai:2003p3625, XZhu, ZhuLiContentAware, nejati2010distortion, katsaggelos, Girod, num_satellite, TVT, Kang:2007p3384, Yan:2006p3580}. 
Following the seminal works on rate control for elastic traffic \cite{Kelly,Low}, there have been several works related to rate control and resource allocation for video transmission in the context of Network Utility Maximization (NUM) frameworks that have considered resource allocation in both wired and wireless networks under different conditions (e.g. \cite{LayeringChiang} and the references therein). The majority of such studies focused on elastic flows whose utility function is known to be continuous and strictly concave. Such an assumption makes the rate control problem convex and thereby tractable for achieving globally optimal solution thanks to dual methods \cite{Low, LayeringChiang}.

On the other hand, stringent QoS requirements of video streams makes them inelastic flow, whose behavior are characterized by nonconcave and often discontinuous utility functions \cite{Ya-Qin_Zhang,Shenker}. This characterization usually eventuates in nonsmooth and nonconvex optimization problems, which are difficult to solve in general through dual methods. There exist several works focusing on rate allocation for such inelastic flows under different scenarios \cite{SLOW2,ChiangInelastic,InelasticWSN,Shroff,Talebi_ICC12,Talebi-Elsevier,
Abbas,Tang}. 
In most developments thus far achieved to handle NUM problems, dual methods have played an important role as a consequence of convexity. To benefit from the developments originally proposed for elastic flows, many researchers aimed to \textit{convexify} the NUM problem or to approximate it by a convex problem. Towards this, in most cases researchers aimed at redefining the notion of the utility or objective so that it yields a convex reformulation;  in some cases, however, the transformed problem is not equivalent to the original problem.   

In SVC streams like other multimedia applications featuring layered encoding schemes, user satisfaction can be considered as having distinct utility levels representing quality indices. This corresponds to the notion of ``discretely adaptive'' utility function for each layer, whose extension yields a \emph{staircase} utility function for SVC streams, as shown in Fig. 1 in solid line \cite{Ya-Qin_Zhang,Zegura}. 

In a previous study \cite{Talebi-Elsevier}, we proposed an analytical model for utility function of SVC-encoded video streams through approximating the ideal staircase utility function to tackle nonsmoothness problem. This approximation was called \emph{Multimodal Sigmoid Approximation}.
Then we redefined approximated utility function through \textit{utility-proportionally fairness} metric which yielded a convex NUM problem. Although the redefinition of utility function in this way was led to a convex formulation, the redefined problem is not equivalent to the original NUM problem. 

\begin{figure}[t]
\begin{center}
\includegraphics[angle=0,scale=.48]{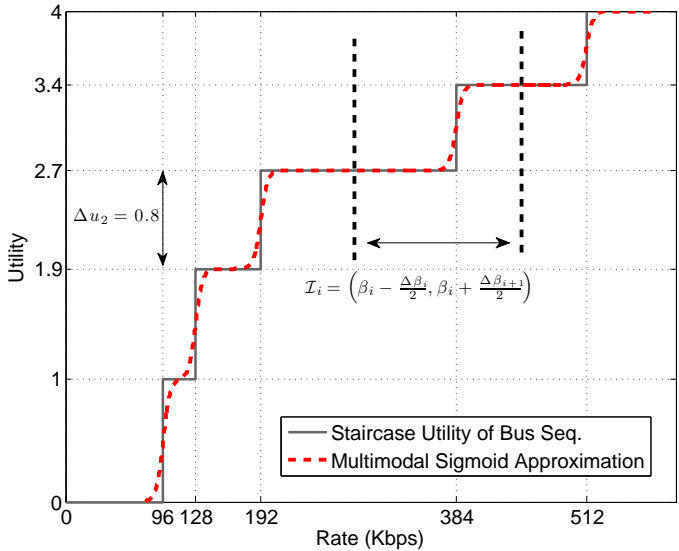}
\end{center}
\caption{Staircase utility model for \emph{Bus} sequence and its Multimodal Sigmoid Approximation}
\label{fig_um}
\end{figure}

Under the above assumption, we believe that this paper makes the following contributions. 
\begin{itemize}
	\item We extend the utility model for SVC streams proposed in \cite{Talebi-Elsevier} such that one can embed Quality-of-Experience (QoE) of the end-user into the notion of utility function. Towards this, QoE indicators can be included as input parameters to yield a QoE-aware rate control scheme for SVC streaming. This way, the \emph{staircase} utility function will possess two degrees of freedom to control preferences of users: rates and QoE indicators. 

	\item As a follow-on work to our previous study \cite{Talebi-GC}, another contribution of this work is to approximate the NUM-based rate control for SVC streams by a D.C. (Difference of Convex) problem \cite{global_opt}. In later sections, using a transformation, we manipulate an approximation of the nonconcave objective function to obtain a strictly concave objective function. These manipulations, however, transform the linear capacity constraints of the original NUM problem into nonconvex ones. Indeed, as we will discuss in details, the transformed capacity constraints admit the notion of a D.C. function. Thus, we achieve a D.C. problem which comprises a strictly concave objective and D.C. constraints. While our analysis here is more rigorous than that of \cite{Talebi-GC}, our developments in this study go further still, where we give sufficient conditions for utility function characterization in terms of QoE indicators to yield strict concavity.

	\item In order to tackle the aforementioned nonconvex D.C. problem, we leverage Sequential Convex Programming (SCP) approach (see e.g., \cite{SCP_EE364,SCP_TranDiehl}). Using SCP approach, one can replace the nonconvex D.C. problem by a sequence of convex programs that try to successively approximate the D.C. problem. Strict concavity of the objective of our D.C. problem guarantees the strict convexity of the aforementioned (approximating) convex programs. This allows us to achieve globally optimal solution of each convex program through solving its dual with gradient projection method \cite{Bert_NLP}. This procedure is the sketch of two rate control algorithms to be discussed in further details in subsequent sections. 
Under mild conditions on the starting point of the algorithms and some parameters, convergence to a KKT point of the D.C. problem is guaranteed. Indeed such a point might be a locally-optimal point of the D.C. problem\footnote{Note that the complexity of D.C. problems is NP-hard. 
It should be noted that the convergence time using centralized methods such as branch and bound (with non-polynomial time complexity guarantee) can be quite long even for problems with relatively small size \cite{global_opt}.}, which is quasi-optimal to the original NUM\footnote{Since we devise rate allocation by solving an approximated version of the network utility in lieu of the original NUM, we refer to this development as quasi-optimal solution to the original NUM problem.}.
\end{itemize} 
Finally, extensive simulation experiments allow us to reach  conclusions regarding the efficacy of the proposed rate control algorithms for SVC streams.  

The rest of the paper is structured as follows. Section \ref{sec:rel_work} reviews some related works. 
Section \ref{sec:system_model} describes the network model and utility function approximation. Section \ref{sec:prob_formulation} is devoted to formulate the underlying NUM problem and its reformulation and approximation as a nonconvex D.C. convex problem. It also describes SCP approach to deal with D.C. problem. Section \ref{sec:opt_solution} investigates the optimal solution to the NUM problem. Section \ref{sec:Algs} presents distributed iterative algorithms for the rate control of SVC streams. Simulation results are given in Section \ref{sec:simulation}. We conclude the paper in Section \ref{sec:conclusion}.

\section{Related Work}
\label{sec:rel_work}
As already mentioned, within the NUM framework, there have been a lot of prior works that propose several protocols and algorithms under different types of traffics, assumptions, and constraints (see \cite{LayeringChiang} and the references therein). In particular, some of existing research on NUM focus on the \textit{inelastic traffic} \cite{SLOW2,ChiangInelastic,Shroff,Talebi_ICC12,InelasticWSN}, that correspond to rate-adaptive transmission scenarios, such as video streaming \cite{Shenker}. The inelasticity will affect the utility function and turns the utility maximization problem into a nonconvex one. Some inelastic flows (e.g. some types of video streams) have been modeled  by sigmoid-like utility functions \cite{Shenker}. We note, however, that such  functions cannot capture the characteristics of SVC streams. 

In our work, the main focus will be on NUM-based rate allocation in SVC transmission scenarios, where the utility function is staircase function. By employing staircase utilities, it is possible to take into account both video characteristics and preferences of users as QoE indicators in rate allocation.

In \cite{Shroff}, the authors adopted sigmoid utility functions and proposed a distributed admission control approach for such utilities, called ``self-regulating'' heuristic. Although the proposed method in \cite{Shroff} can be extended for our approximated utility model, but rate allocation in our solution will employ no heuristics for preventing divergence of the algorithm.   
Hande \emph{et al.} in \cite{ChiangInelastic} investigated the optimality conditions for distributed iterative dual-based algorithm to converge to globally optimal point despite using nonconcave utility functions. Although deriving the sufficient and necessary conditions for such a nonconvex problem is valuable, this work does not pragmatically solve the inelastic rate allocation problem. 
Sehati \emph{et al.} in \cite{Talebi_ICC12}, addressed rate allocation  using the NUM framework for streaming traffic whose characteristic can be captured by a specific sigmoidal-like utility function called S-curve function. They employed Sequential Convex Programming approach to devise a distributed rate control algorithm as a sub-optimal solution to the nonconvex NUM problem. 
In \cite{SLOW2}, through \textit{Utility-proportional fairness} metric, a modified version of utility function is introduced that is appropriate for heterogeneous networks carrying both elastic and inelastic flows.

In \cite{Abbas} and \cite{Tang}, using particle swarm optimization (PSO) \cite{pso}, the authors proposed two solutions for maximizing network utility of both elastic and inelastic flows. PSO is an evolutionary algorithm that can be used for discontinuous, non-convex and nonlinear problems such as ours. But, the main problem is that this algorithm is centralized and is not desirable in networking scenarios. In addition, due to randomized nature of this algorithm, convergence cannot be guaranteed.

Some research studies adopted the rate-distortion function as the utility function for rate control of SVC streams \cite{nejati2010distortion,Girod,TVT,katsaggelos,ZhuLiContentAware,num_satellite}. The authors of \cite{katsaggelos} chose PSNR function as utility function and proposed a distributed approach for resource allocation in video streaming scenarios. In \cite{ZhuLiContentAware}, a content-aware distortion-fair networking framework with joint video source adaptation and network resource allocation is developed. 
The main difference between these works with ours is that employing distortion or PSNR as objective function captures the content characteristics of video transmission, while staircase utility function is more general and captures the traffic characteristics as well as preferences of users in  SVC streaming scenarios.


\subsection{Quality of Experience}
In this subsection we review some facts about Quality of Experience (QoE). QoE is defined in \cite{QoE2} as \emph{``a multi-dimensional construct of perceptions and behaviors of a user, which represents his/her emotional, cognitive, and behavioral responses, both subjective and objective, while using a system''}. While traditional well-established notions of Quality of Service (QoS) are considered as network-related metrics in terms of bandwidth, delay, jitter, and packet loss, QoE is a user-related metric. In the case of multimedia streaming, from user's point of view, QoS indicators are not sufficient to get an accurate idea about the final acceptability of the received content. On the other hand, the main focus of QoE is on the overall experience of user. Thus, some recent studies put effort on replacement of QoS by QoE, since the latter is more correlated to the user's preferences.  

In our NUM framework, input parameters of the problem (staircase function parameters) can be interpreted as QoE indicators of users. In particular, the underlying \emph{staircase} utility function we consider is QoE-aware. As two parameters that define the \emph{staircase} utility, rate points and quality indices can capture user's preferences in terms of QoE indicators. We defer more discussions to Section \ref{sec:um}.

\section{System Model}
\label{sec:system_model}
\subsection{Network Model}
We consider a network comprising a set of sources denoted by \mbox{$\mathcal S=\{1,\dots,S\}$} that share a set of logical links denoted by \mbox{$\mathcal L=\{1,\dots,L\}$}. We denote by $\boldsymbol c=(c_l,l\in\mathcal L)$ the link capacity vector where $c_l$ is the capacity of link $l$ in bps.  
Let $x_s\in\mathcal X_s=[m_s, M_s]$ be the rate of source $s$ in bps and $m_s$ and $M_s$ be the minimum and the maximum rate of source $s$, respectively. Also let \mbox{$\boldsymbol{\mathcal X}=\varotimes_{s=1}^S\mathcal X_s$} denote the Cartesian product of $\mathcal X_s, s=1,\dots,S$.

We assume that each logical source transmits one video session and thus we will use terms `session' and `source', interchangeably.  
We focus on static routing and to represent the links that source $s$ passes through, we define the routing matrix as $\mathbf{R}=[R_{ls}]_{L\times S}$, where $R_{ls}$ is defined as 
\begin{equation}
R_{ls}=\left \{ \begin{array}{ll}
1 & \qquad \textrm{if session $s$ passes through link $l$}\\
0 & \qquad \textrm{otherwise}\\	
\end{array} \right.
\end{equation}

In order to simplify the analysis and not relying on any particular packet scheduling and flow control scheme, rate control is ideally accommodated by congestion in links. Thus, a rate allocation vector $\boldsymbol x=(x_s,s\in\mathcal S)$ is feasible if and only if $\boldsymbol x\in\boldsymbol{\mathcal X}$ and flow of each link is less than its capacity, i.e. 
$$\sum_{s=1}^S R_{ls}x_s\leq c_l;\quad l\in\mathcal L.$$

Formally speaking, we define the feasible rate region $\boldsymbol{\mathcal D}$ as the following polyhedron 
\begin{equation}
\label{eq:rate_region}
\boldsymbol{\mathcal D}=\left\{\boldsymbol x\in\boldsymbol{\mathcal X}\big|\mathbf R\boldsymbol x\leq \boldsymbol c\right\}.
\end{equation}

\subsection{Utility Model \label{sec:um}}
We suppose that source $s$, when submitting its video session at rate $x_s$, attains a utility $U_s(x_s)$ which captures the level of quality perceived by user. In this paper, we assume that video session $s$ is encoded in compliance with SVC standard and this subsection is devoted to model the utility function for SVC-encoded streams.  

For video sequences encoded in compliance with SVC standard, rate allocation is ideally limited to distinct levels of quality. This means that the utility function is increased only when a higher layer can be delivered due to increase in the available bandwidth. Thus, the ideal utility function for such sequences is characterized using a \emph{staircase} function, which is shown for \emph{Bus} sequence in Fig. 1 in solid line. 

As discussed in Section \ref{sec:intro}, for rate-adaptive multimedia streams, NUM problems are usually nonconvex and nonsmooth. In particular, for SVC streams whose ideal utility function is staircase function, both nonconvexity and nonsmoothness issues exist.
In our previous studies \cite{Talebi-Elsevier,Talebi-GC}, we proposed a smoothed approximation of the ideal utility function for SVC streams, referred to as \emph{multimodal sigmoid approximation}. This approximation is shown in Fig. 1 in dashed line. In this study, we consider a more accurate utility function as the approximation to staircase utility function that takes into account a more general quality model for different layers.

In what follows, we briefly describe the construction of utility approximation. We consider the ideal staircase utility function $U^{\mathrm{ideal}}(x)$ in general form defined by a quality index sequence $\{u_i\}_{i=0,\dots,N}$ and a rate index sequence $\{\beta_i\}_{i=0,\dots,N}$ such that 
\begin{equation}
U^{\mathrm{ideal}}(x)=u_{i+1};\quad x\in(\beta_i,\beta_{i+1}],\quad i=0,\dots,N-1
\end{equation}

Rate points in utility function reflect the layer changing points and thus are defined on a per sequence basis. Hence, for each video sequence we would have a specific utility function. 
Therefore, rate allocation will be influenced by users' QoE as a result of having different video sequences. 
On the other hand, according to \cite{QoE}, QoE is defined as the overall acceptability of received content. Beside visual quality of the content, as one indicator for acceptability judgment, some other parameters have also influence on QoE. One such parameter is the type of receiving platform of the end user. Here we do not restrict the rate allocation to any specific receiving platform. As such, users can be considered as having different priorities for video resolutions. While higher resolutions (i.e. higher spatial layers) are completely satisfactory for users with high resolution screens, it is not appropriate for those with handheld devices. Thus, different levels of utility function at each rate point (defined using quality index sequence) can vary based on the end user's receiving platform. 

\subsection{Utility Approximation}
Let $U(x)$, $x\in\mathcal X$ denote the smoothed utility function. We assume that the $i$th transition in $U(x)$, i.e. transition of utility function from quality index $u_i$ to $u_{i+1}$, occurs at rate index $x=\beta_i$. We refer to this transition as the $i$th step. Thus, we divide the domain $\mathcal X$ into $N$ intervals $\mathcal I_0, \mathcal I_1, \dots, \mathcal I_{N-1}$ with non-overlapping interiors\footnote{We denote by $\mathbf{int}\textrm{ }\mathcal Z$ the interior of set $\mathcal Z$.} such that \mbox{$\beta_i\in\mathbf{int}\textrm{ }\mathcal I_i,i=0,\dots,N-1$}. Focusing on two contiguous steps, say $i$ and $i+1$, we define the lower and upper boundary points of $\mathcal I_i$ to be the middle points of $\left[\beta_{i-1},\beta_{i}\right]$ and $[\beta_i,\beta_{i+1}]$ intervals, respectively. Now we can write 
\begin{equation}
\mathcal I_i=\left[\beta_{i}-\frac{\Delta\beta_{i}}{2},\beta_{i}+\frac{\Delta\beta_{i+1}}{2}\right]
\end{equation}
where
$$\Delta\beta_{i}\triangleq\beta_i-\beta_{i-1}.$$
Then, the underlying idea is to define $U(x)$ as a smoothed version of $U^{\mathrm{ideal}}(x)$ by $N$ properly defined sigmoid functions. Towards this, for $x\in\mathcal I_i$ we define $U(x)$ as follows 
\begin{equation}
U(x)=\left(u_{i+1}-u_i\right)F(x,\alpha,\beta_i)+u_i; \qquad x\in\mathcal I_i
\end{equation} 
where $F(x,\alpha,\beta_i)$ is the sigmoid function with parameters $(\alpha,\beta_i)$ and is given by
\begin{equation}
F(x,\alpha,\beta_i)=\frac{1}{1+e^{-\alpha(x-\beta_i)}}
\end{equation} 
It's easy to show that $x=\beta_i$ is the inflection point of the sigmoid function, and given $\beta_i$, $\alpha$ determines how good the sigmoid function approximates the step curve. 
Then, the approximated utility $U(x)$ can be expressed by
\begin{equation}
\label{eq:Multimodal_sigmoid_approximate_U}
U(x)=\left \{\begin{array}{ll}
(u_{1}-u_{0})F(x,\alpha,\beta_0)+u_0 & x\in\mathcal I_0\\
\vdots\\
\left(u_{i+1}-u_i\right)F(x,\alpha,\beta_i)+u_i & x\in\mathcal I_i\\
\vdots\\
\left(u_{N}-u_{N-1}\right)F(x,\alpha,\beta_{N-1})+u_{N-1} & x\in\mathcal I_{N-1}
\end{array} \right.
\end{equation}

Continuity of $U(x)$ is relying on the choice of sigmoid parameters $\alpha$ and $\{\beta_{i}\}_{i=0,\dots,N}$. If they are chosen such that \mbox{$G_i\triangleq \exp{\left(\alpha\frac{\Delta\beta_i}{2}\right)},i=1,\dots,N$} are sufficiently large, the discontinuity gap between contiguous sigmoids vanishes, thus making  $U(x)$ continuous and differentiable, as shown in Fig. 1. 

In this paper, we assume that session $s$ has $N_s$ layers. Then, representing its quality index sequence and rate index sequence, respectively by $\{u_{si}\}_{i=0,\dots,N_s}$ and $\{\beta_{si}\}_{i=0,\dots,N_s}$, we express utility function $U_s(x_s)$ for video session $s$ by
\begin{equation}
\label{eq:Us_def}
U_s(x_s)=U(x_s,\alpha_s,\{\beta_{si}\})
\end{equation} 

\section{Problem Formulation and Approximation}
\label{sec:prob_formulation}
We model the rate control algorithm for SVC-encoded streams as the solution to a NUM problem. The objective of such a NUM problem is the aggregate utility of all sources. Nondifferentiability of the ideal utility function can be remedied by employing the multimodal sigmoid approximation introduced above. Therefore, we consider the following NUM problem
\begin{eqnarray}
\label{eq:std_NUM}
\max_{\boldsymbol x\in\boldsymbol{\mathcal D}} \sum_{s\in\mathcal S}w_sU_s(x_s)
\end{eqnarray}
where $w_s,s=1,\dots,S$ are normalized positive weights such that $\sum_s w_s=1$. 
Problem (\ref{eq:std_NUM}) is nonconvex because of nonconcavity of $U_s$. 

\subsection{Objective Function Approximation}
In order to come up with a more amenable formulation, we consider the following optimization problem
\begin{equation}
\label{eq:opt_apprx}
\max_{\boldsymbol x\in\boldsymbol{\mathcal D}} \sum_{s\in\mathcal S} w_s \log U_s(x_s).
\end{equation} 
The following lemma states that problem (\ref{eq:opt_apprx}) approximates problem (\ref{eq:std_NUM}). 

\begin{myLemma}
The optimal solution of problem (\ref{eq:opt_apprx}) gives a lower bound to the optimal solution of problem (\ref{eq:std_NUM}).
\end{myLemma}

\begin{proof}
Taking the logarithm of the objective of (\ref{eq:std_NUM}) yields
\begin{equation}
\label{eq:log-opt}
\max_{\boldsymbol x\in\boldsymbol{\mathcal D}} \log\left(\sum_s w_sU_s(x_s)\right)
\end{equation}
Since $\log(.)$ function is monotonically increasing, maximizing (\ref{eq:log-opt}) is equivalent to maximizing (\ref{eq:std_NUM}), and thereby problems (\ref{eq:log-opt}) and (\ref{eq:std_NUM}) are equivalent \cite{Boyd}. On the other hand, $\log(.)$ is a concave function; hence for $z_s>0, s\in\mathcal S$, we have 
\begin{equation}
\label{eq:jensen_inequality}
\log\left(\sum_s w_s z_s\right) \geq \sum_s w_s\log z_s
\end{equation}
provided that $w_s\geq 0$ and $\sum_s w_s=1$. 
Then, provided that $U_s(x_s)>0$, using (\ref{eq:jensen_inequality}) we get
\begin{equation}
\label{eq:jensen_Us}
\log\left(\sum_s w_s U_s(x_s)\right)\geq \sum_s w_s\log U_s(x_s)
\end{equation}
This tells us that the R.H.S of (\ref{eq:jensen_Us}) is the lower bound of the objective of (\ref{eq:log-opt}). As (\ref{eq:jensen_Us}) holds for all feasible points $\boldsymbol x\in\boldsymbol{\mathcal D}$, at the optimal point, (\ref{eq:log-opt}) is lower bounded by (\ref{eq:opt_apprx}). Finally, equivalence of (\ref{eq:std_NUM}) and (\ref{eq:log-opt}) completes the proof.
\end{proof}

It's straightforward to confirm that the objective of (\ref{eq:opt_apprx}) is nonconcave. In order to come up with a concave objective, we use the following transformation 

\begin{equation}
\label{eq:exp_transform}
\tilde x_s=e^{\alpha_s x_s}
\end{equation}
Substituting $x_s=\frac{1}{\alpha_s}\log{\tilde x_s}$, we reformulate the new objective as
\begin{eqnarray}
\tilde U_s(\tilde x_s)&\triangleq&\log U_s(x_s)\nonumber\\
								&=&\log U_s\left(\frac{1}{\alpha_s}\log \tilde x_s\right);\nonumber\\
								&&\qquad\qquad\qquad \tilde x_s\in\mathcal{\tilde X}_s\triangleq\left[e^{\alpha_s m_s},e^{\alpha_s M_s}\right].\nonumber
\end{eqnarray}
Concentrating on the $i$th interval, by further simplifying the above we get
\begin{eqnarray}
\tilde U_s(\tilde x_s)&=&\log U_s\left(\frac{1}{\alpha_s}\log \tilde x_s\right)\nonumber\\
&=&\log \left[\Delta u_{s(i+1)}F\left(\frac{1}{\alpha_s}\log\tilde x_s,\alpha_s,\beta_{si}\right)+u_{si}\right]\nonumber\\
&=& \log\left(\frac{\Delta u_{s(i+1)}}{1+e^{-\alpha_s(\frac{1}{\alpha_s}\log\tilde x_s-\beta_{si})}}+u_{si}\right)\nonumber\\
\label{eq:transformed_utility_explicit}
&=& \log\left(\frac{\Delta u_{s(i+1)}\tilde x_s}{\tilde x_s+e^{\alpha_s\beta_{si}}}+u_{si}\right);\quad \tilde x_s\in\mathcal{\tilde I}_{si}
\end{eqnarray}
where $\mathcal{\tilde I}_{si}$ is the image of $\mathcal I_{si}$ under mapping (\ref{eq:exp_transform}).
The following lemma determines the conditions under which the transformed utility function $\tilde U_s(\tilde x_s)$ is strictly concave over $\tilde x_s\in\mathcal{\tilde X}_s$.

\begin{myLemma}
\label{concave_Lemma}
Provided that the quality index sequence $\{u_{i}\}_{i=0,\dots,N}$ is monotonically increasing and strictly concave\footnote{A sequence $\{a_i\}_{i=0,\dots,N}$ is said to be concave if the inequality \mbox{$a_{i}\geq \left(a_{i-1}+a_{i+1}\right)/2$} holds for every $a_i, i=2,\dots,N-1$. This sequence is said to be strictly concave if the inequality holds strictly.} and $G_i$'s are sufficiently large, i.e.
\begin{enumerate}
\item[\textbf{C1:}] $u_{i+1}> u_i;\quad i=1,\dots,N-1$ 
\item[\textbf{C2:}] $u_{i+1}-u_i>u_{i+2}-u_{i+1};\quad i=1,\dots,N-2$ 
\item[\textbf{C3:}] $G_i\triangleq \exp{\left(\alpha\frac{\Delta\beta_i}{2}\right)}\gg 1;\quad i=1,\dots,N$
\end{enumerate}
then the transformed utility function $\tilde U(\tilde x)$ is strictly concave.
\end{myLemma}

\begin{proof}
See Appendix I.
\end{proof}


We reformulate problem (\ref{eq:opt_apprx}) by substituting transformed utility functions as 
\begin{equation}
\label{eq:opt_transformed}
\max_{\boldsymbol{\tilde x}\in\boldsymbol{\tilde{\mathcal D}}} \sum_{s\in\mathcal S} w_s\tilde U_s(\tilde x_s)
\end{equation} 
where $\boldsymbol{\tilde x}=(\tilde x_s,s\in\mathcal S)$ represents the transformed rate vector and $\boldsymbol{\tilde{\mathcal D}}$ is the feasible region of the transformed problem which is defined by
\begin{equation}
\label{eq:rate_region}
\boldsymbol{\tilde{\mathcal D}}=\left\{\boldsymbol{\tilde x}\in\boldsymbol{\mathcal{\tilde X}}\biggr|\sum_s \frac{R_{ls}}{\alpha_s}\log \tilde x_s \leq c_l, l\in\mathcal L\right\}
\end{equation}
where $\boldsymbol{\mathcal{\tilde X}}=\varotimes_{s=1}^S\mathcal{\tilde X}_s$.
Problems (\ref{eq:opt_apprx}) and (\ref{eq:opt_transformed}) are equivalent as (\ref{eq:exp_transform}) is monotonically increasing and its domain covers $\boldsymbol{\mathcal{D}}$ \cite{Boyd}. 

\subsection{Sequential Convex Programming (SCP) Approach}
Unfortunately the feasible region of the transformed problem $\boldsymbol{\tilde{\mathcal D}}$ is a nonconvex set. To show this, we consider $\boldsymbol{\tilde{\mathcal D}}$ as the intersection of $0$-sublevel sets\footnote{For a function $f:\mathbb R^n\rightarrow\mathbb R$, the associated $\alpha$-sublevel set is a subset of the domain whose elements yield $f(x)\leq \alpha$, or more formally \mbox{$C_\alpha=\{x\in\mathbf{dom}\textrm{ }f\big|f(x)\leq\alpha\}$} \cite{Boyd}.} of $L$ functions defined by
$$g_l(\boldsymbol{\tilde x})\triangleq\sum_s \frac{R_{ls}}{\alpha_s}\log \tilde x_s-c_l;\quad l\in\mathcal L$$
The $0$-sublevel set of $g_l(.)$ is convex if and only if $g_l(.)$ is quasiconvex. The function $\log\tilde x_s$ is quasiconvex, however, a positively weighted sum of quasiconvex functions is not necessarily quasiconvex \cite{Boyd}. Indeed, simple algebraic operations can show the failure of quasiconvexity of $g_l(.)$. Therefore, $g_l(.)$ is not a quasiconvex function and thereby $\boldsymbol{\tilde{\mathcal D}}$ is nonconvex. 
Nonconvexity of $\boldsymbol{\tilde{\mathcal D}}$ makes problem (\ref{eq:opt_transformed}) nonconvex, too. We note that $g_l(.)$ can be written as the difference of $-c_l$ and $-\sum_s w_s\log \tilde x_s$, i.e. as the difference of two convex functions. Thus it is a \emph{Difference of Convex (D.C.)} function \cite{global_opt}.  
In particular, in this specific case where $g_l(.)$ is a concave function, the constraint $g_l(\boldsymbol{\tilde x})\leq 0$ is also referred to as a \emph{reverse-convex constraint} as $g_l(\boldsymbol{\tilde x})\geq 0$ corresponds to a convex constraint. Reverse-convex constraints are special cases of \emph{Difference of Convex (D.C.) constraints} \cite{global_opt}.

In order to tackle such D.C. constraints above, we use the \emph{Sequential Convex Programming (SCP)} approach (see e.g. \cite{SCP_EE364,SCP_TranDiehl,Talebi_ICC12}). In SCP approach the original nonconvex problem is approximated by a series of convex problems, where each convex problem is constructed as an approximate to the nonconvex problem in a feasible point. Since nonconvexity of our problem is due to nonconvexity of reverse-convex constraints, employing SCP approach for that is equivalent to approximating them by a set of convexified constraints successively. This has also been referred to as \emph{Successive Approximation Technique} in some recent works, e.g. \cite{Evans_TSP}. 

The SCP approach works as follows. Given the auxiliary feasible rate vector $\boldsymbol{z}\in\boldsymbol{\tilde{\mathcal D}}$, the L.H.S of each reverse-convex constraint $g_l(\boldsymbol{\tilde x})\leq 0$ is replaced by its convex majorant, which is the first order Taylor approximation around $\boldsymbol{z}$, denoted by $\hat g_l(\boldsymbol{\tilde x},\boldsymbol{z})$, as follows
\begin{eqnarray}
\hat g_l(\boldsymbol{z},\boldsymbol{\tilde x})&=&g_l(\boldsymbol{z})+\nabla g_l(\boldsymbol{z})^T(\boldsymbol{\tilde x}-\boldsymbol{z})\nonumber\\
&=&\sum_s\frac{R_{ls}}{\alpha_s}\log z_s-c_l+\sum_s\frac{R_{ls}}{\alpha_s}\left(\frac{\tilde x_s-z_s}{z_s}\right).\nonumber
\end{eqnarray}

Since $g_l(.)$ is differentiable, $\nabla g_l$ exists at auxiliary vector $\boldsymbol{z}\in\boldsymbol{\tilde{\mathcal D}}$. It's easy to verify that $\hat g_l(\boldsymbol{\tilde x},\boldsymbol{z})$ is affine in $\boldsymbol{\tilde x}$ and thereby $\hat g_l(.)$ is convex. Thus, the constraint $\hat g_l(\boldsymbol{\tilde x},\boldsymbol{z})\leq 0$ represents a convex constraint.

Indeed, the intersection of $L$ constraints $\hat g_l(\boldsymbol{\tilde x},\boldsymbol{z})\leq 0, \forall l$ is contained in the nonconvex feasible region made by D.C. constraints, and thereby plays the role of an approximate to the nonconvex feasible region.  
Using the approximated feasible region, we obtain a convex problem that can provide arbitrarily good approximation to problem with D.C. constraints. Such a good approximate, however, essentially relies on the knowledge of a `good' feasible vector $\boldsymbol z$, i.e. a feasible vector such that the optimal point of the problem with D.C. constraints lies in the approximated feasible region. Such a vector might not be known a priori. Thus, we have to successively approximate the feasible region so as to obtain such a `good' rate vector.

To this end, we consider the following iterative setting (Algorithm 1). Let $\boldsymbol{\tilde x}^{(0)}$ be an arbitrary feasible point. Then, at the $k$th iterate, we find 
$$\boldsymbol{\tilde x}^{(k+1)}\in\{\boldsymbol z\big|\boldsymbol z\quad \textrm{solves}\quad \textsf{P}(\boldsymbol{\tilde x}^{(k)})\},$$ 
where 
\begin{eqnarray}
\label{eq:opt_transformed_affine_constr}
\textsf{P}(\boldsymbol{\tilde x}^{(k)}):&& \max_{\boldsymbol{\tilde x}\in\boldsymbol{\tilde{\mathcal X}}} \sum_{s\in\mathcal S} w_s\tilde U_s(\tilde x_s)\\
&&\textrm{subject to:}\nonumber\\
&&\sum_s\frac{R_{ls}}{\alpha_s}\left(\log \tilde x^{(k)}_s+\frac{\tilde x_s-\tilde x^{(k)}_s}{\tilde x^{(k)}_s}\right)\leq c_l;\quad l\in\mathcal L.\nonumber\\
\label{eq:const_affined}
\end{eqnarray}


\begin{myTheo}
Suppose $\boldsymbol{\tilde x}^{(0)}\in\boldsymbol{\tilde{\mathcal D}}$. Provided that transformed utility functions $\tilde U_s(\tilde x_s),s\in\mathcal S$ satisfy the conditions in Lemma \ref{concave_Lemma}, then for each $k$, the problem $\textsf{P}(\boldsymbol{\tilde x}^{(k)})$ defined by   (\ref{eq:opt_transformed_affine_constr})-(\ref{eq:const_affined}) is strictly convex and admits a unique maximizer.
\end{myTheo}

\begin{proof}
According to Lemma \ref{concave_Lemma}, provided that conditions \textbf{C1}-\textbf{C3} hold for all sources, $\tilde U_s(\tilde x_s)$ is strictly concave. As the objective of problem (\ref{eq:opt_transformed_affine_constr})-(\ref{eq:const_affined}) is a nonnegative and nonzero weighted sum of strictly concave functions, it is strictly concave too.
 
Constraints (\ref{eq:const_affined}) are affine functions and thereby are convex too. Therefore, we deduce that the optimization problem (\ref{eq:opt_transformed_affine_constr})-(\ref{eq:const_affined}) is strictly convex \cite{Boyd}. Since the feasible set is compact, at least an optimal solution exists. Strict convexity of the problem guarantees that the optimal solution is unique.
\end{proof}

\footnotesize
\begin{center}
\begin{tabular}[!th]{l}\
\ \\
\textbf{Algorithm 1.} Sequential Convex Programming (SCP) 
Procedure\\
\hline
\ \\
\textbf{Initialization.}\\
\quad - Choose an arbitrary feasible vector $\boldsymbol{\tilde x}^{(0)}\in\boldsymbol{\tilde{\mathcal D}}$.\\
\quad - Establish problem $\textsf{P}(\boldsymbol{\tilde x}^{(0)})$.\\
\quad - Set $k=0$.\\
\ \\
\textbf{Main Loop.}\\
\quad - At $k$th iterate, obtain $\boldsymbol{\tilde x}^{(k+1)}$ by solving $\textsf{P}(\boldsymbol{\tilde x}^{(k)})$.\\
\quad - Set $k\leftarrow k+1$ and repeat.\\
\ \\
\hline
\ \\
\end{tabular}
\end{center}
\normalsize

We defer solving the optimization problem until the next section. 

\section{Optimal Solution}
\label{sec:opt_solution}
In this section, we derive an iterative solution to solve problem (\ref{eq:opt_transformed_affine_constr})-(\ref{eq:const_affined}). 


\subsection{Primal Optimality}
Let $L_k(\boldsymbol{\tilde x},\boldsymbol\mu)$ be the Lagrangian associated to problem $\textsf{P}(\boldsymbol{\tilde x}^{(k)})$ defined by (\ref{eq:opt_transformed_affine_constr})-(\ref{eq:const_affined}). Then
\begin{equation}
\label{eq:Lagrangian}
L_k(\boldsymbol{\tilde x},\boldsymbol\mu)=\sum_s w_s\tilde U_s(\tilde x_s)-\sum_l \mu_l\hat g_l(\boldsymbol{\tilde x},\boldsymbol{\tilde x}^{(k)})
\end{equation}
where $\mu_l$ is the positive Lagrange multiplier associated to approximated capacity constraint (\ref{eq:const_affined}) for link $l$ and $\boldsymbol \mu=(\mu_l, l\in\mathcal{L})$ is a vector of Lagrange multipliers. 

According to KKT theorem, the optimal solution of problem $\textsf{P}(\boldsymbol{\tilde x}^{(k)})$ is the stationary point of the Lagrangian $L_k$ or equivalently its maximizer \cite{Boyd}. As introduced in the previous section, $\boldsymbol{\tilde x}^{(k+1)}$ solves $\textsf{P}(\boldsymbol{\tilde x}^{(k)})$. Hence, it is the maximizer of the corresponding Lagrangian, or more formally
$$\boldsymbol{\tilde x}^{(k+1)}(\boldsymbol\mu)=\arg\max_{\boldsymbol{\tilde x}\in\boldsymbol{\mathcal{\tilde X}}} L_k(\boldsymbol{\tilde x},\boldsymbol\mu)$$ 

\begin{myTheo}
\label{theo:x_opt}
Given dual variable vector $\boldsymbol\mu\in\mathbb R_+^L$, the unique maximizer of $L_k$, i.e. $\boldsymbol{\tilde x}^{(k+1)}(\boldsymbol\mu)$ is given in (\ref{eq:tildexs_opt}), and $\tilde x^{(k+1)}_s(\boldsymbol\mu)$ belongs to $\mathcal{\tilde I}_{si^{(k+1)}_s}$ where $i^{(k+1)}_s$ is the solution to the following inequality
\begin{figure*}
\begin{equation}
\label{eq:tildexs_opt}
\tilde x_s^{(k+1)}(\boldsymbol\mu)=\frac{A_{si^{(k+1)}_{s}}}{2u_{s(i^{(k+1)}_{s}+1)}}\left(\Delta u_{s(i^{(k+1)}_s+1)}\sqrt{1+\frac{4w_s\alpha_s\tilde x^{(k)}_su_{s(i^{(k+1)}_s+1)}}{\Delta u_{s(i^{(k+1)}_s+1)}\mu^sA_{si^{(k+1)}_s}}}-u_{si^{(k+1)}_{s}}-u_{s(i^{(k+1)}_{s}+1)}\right)
\end{equation}
\hrule
\hrulefill
\vspace*{4pt}
\end{figure*}
\begin{equation}
\label{eq:opt_level}
L_{si^{(k+1)}_s}\leq\frac{\mu^s}{\tilde x^{(k)}_s}\leq U_{si^{(k+1)}_s}
\end{equation}
and
\begin{equation}
\label{eq:coefficient_1}
\mu^s\triangleq\sum_l R_{ls}\mu_l
\end{equation}
and $L_{si^{(k+1)}_s}$ and $U_{si^{(k+1)}_s}$ are given in (\ref{eq:coefficient_4}) and (\ref{eq:coefficient_5}), respectively.
\begin{figure*}
\begin{eqnarray}
\label{eq:coefficient_4}
L_{si^{(k+1)}_s}&=&\frac{w_s\alpha_s\Delta u_{s(i^{(k+1)}_s+1)}}{A_{si^{(k+1)}_{s}}\left(G_{s(i^{(k+1)}_s+1)}+1\right)\left(u_{s(i^{(k+1)}_{s}+1)}G_{s(i^{(k+1)}_s+1)}+u_{si^{(k+1)}_{s}}\right)}\\
&&\nonumber\\
\label{eq:coefficient_5}
U_{si^{(k+1)}_s}&=&\frac{w_s\alpha_s\Delta u_{s(i^{(k+1)}_s+1)}G_{si^{(k+1)}_s}^2}{A_{si^{(k+1)}_{s}}\left(G_{si^{(k+1)}_s}+1\right)\left(u_{s(i^{(k+1)}_{s}+1)}+G_{si^{(k+1)}_s}u_{si^{(k+1)}_{s}}\right)}\\
&&\nonumber\\
\textrm{where}\qquad A_{si^{(k+1)}_{s}}&=&\exp(\alpha_s\beta_{si^{(k+1)}_{s}}) \qquad\textrm{and}\qquad 
B_{si^{(k+1)}_{s}}=\frac{u_{i^{(k+1)}_{s}}}{u_{i^{(k+1)}_{s}+1}}\exp(\alpha_s\beta_{si^{(k+1)}_{s}}).
\end{eqnarray}
\hrule
\hrulefill
\vspace*{4pt}
\end{figure*}
\end{myTheo}

\begin{proof}
See Appendix II.
\end{proof}

Optimal source rates can be simply obtained from (\ref{eq:tildexs_opt}) by taking the inverse transformation of (\ref{eq:exp_transform}) as follows
\begin{equation}
\label{eq:xs_opt}
x_s^{(k+1)}(\boldsymbol\mu)=\bigg[\frac{1}{\alpha_s}\log \tilde x_s^{(k+1)}(\boldsymbol\mu)\bigg]_{\mathcal X_s}
\end{equation}
where $[.]_{\mathcal X_s}$ is the projection operator onto $\mathcal X_s$.

\subsection{Dual Optimality}
Theorem \ref{theo:x_opt} gives the optimal solution to the $k$th problem $\textsf{P}(\boldsymbol{\tilde x}^{(k)})$ as a function of Lagrange multiplier vector $\boldsymbol\mu$. To obtain the optimal Lagrange multiplier vector, denoted by $\boldsymbol\mu^*$, one has to solve dual problem associated to problem (\ref{eq:opt_transformed_affine_constr})-(\ref{eq:const_affined}), which is given by \cite{Boyd}:
\begin{equation}
\label{eq:dual_prb}
\textsf{D}_k:\quad\min_{\boldsymbol\mu\geq \mathbf 0} \left\{h_k(\boldsymbol\mu)\triangleq\max_{\boldsymbol{\tilde x}\in \boldsymbol{\tilde{\mathcal X}}} L_k(\boldsymbol{\tilde x},\boldsymbol\mu)\right\}
\end{equation}
where $h_k(\boldsymbol\mu)$ is the dual function associated to problem $\textsf{P}(\boldsymbol{\tilde x}^{(k)})$ and based on this terminology, $\boldsymbol\mu$ is referred to as the vector of dual variables, too.

In view of Theorem \ref{theo:x_opt}, the dual function can be expressed as $h_k(\boldsymbol\mu)=L_k(\boldsymbol{\tilde x}^{(k+1)}(\boldsymbol\mu),\boldsymbol\mu)$. Solving dual problem in closed form is usually impossible. Instead, one can benefit from iterative methods to achieve the solution. 
Due to strict convexity of problem (\ref{eq:opt_transformed_affine_constr})-(\ref{eq:const_affined}), dual function $h_k(\boldsymbol\mu)$ is continuously differentiable over $\mathbb R^L_+$  whose derivatives, by Danskin's Theorem, is characterized by \cite{Bert_NLP}:
\begin{equation}
\label{eq:dual_derivative}
\frac{\partial h_k(\boldsymbol\mu)}{\partial\mu_l}=c_l-\sum_s \frac{R_{ls}}{\alpha_s}\left(\log \tilde x^{(k)}_s+\frac{\tilde x_s-\tilde x^{(k)}_s}{\tilde x^{(k)}_s}\right)
\end{equation}
Due to differentiability of dual function, we solve dual problem (\ref{eq:dual_prb}) by \textit{gradient projection algorithm} \cite{Bert_NLP}.

\begin{figure*}
\begin{eqnarray}
\label{eq:dual-update}
\mu_l^{(t+1)}=\bigg[\mu_l^{(t)}-\gamma_k\frac{\partial h_k(\boldsymbol\mu^{(t)})}{\partial\mu_l}\bigg]^+
=\bigg[\mu_l^{(t)}-\gamma_k\left\{c_l-\sum_s \frac{R_{ls}}{\alpha_s}\left(\log \tilde x_s^{(k)}+\frac{v^{(t)}_s-\tilde x^{(k)}_s}{\tilde x^{(k)}_s}\right)\right\}\bigg]^+\nonumber\\
\end{eqnarray}
\hrule
\hrulefill
\vspace*{4pt}
\end{figure*}

Let $\boldsymbol{v}^{(0)}=\boldsymbol{\tilde x}^{(k)}$ denote the initial (primal) point to solve problem $\textsf{D}_k$. Then, the dual variable update equation at $t$th iterate for solving $\textsf{D}_k$ is given in (\ref{eq:dual-update}) in which  $\gamma_k$ is a sufficiently small constant step size  properly chosen for problem $\textsf{D}_k$ and \mbox{$[z]^+=\max(z,0)$}. 

We defer the algorithmic description of this iterative procedure until the next section. 

\section{Rate Control Algorithms for SVC Streams}
\label{sec:Algs}
Motivated by the iterative solution to problem (\ref{eq:opt_transformed_affine_constr})-(\ref{eq:const_affined}) obtained above as well as SCP procedure outlined as Algorithm 1, we propose two rate control algorithms for SVC streams. 
These algorithms are governed by optimal rate equations (\ref{eq:tildexs_opt}), (\ref{eq:xs_opt}), and (\ref{eq:vs_iteration_t}), and optimal index inequalities (\ref{eq:opt_level})-(\ref{eq:coefficient_5}), and dual variable update equations (\ref{eq:dual-update}).

\subsection{Algorithm 2: The Two-Tier Algorithm}
The first rate control algorithm, listed as Algorithm 2, is a two-tier algorithm as it possesses two kinds of iteration: outer and inner iterations. The outer iteration corresponds to the iterations of SCP procedure (Algorithm 1). The inner iteration corresponds to the iterations required to solve the dual of each convex program at each outer iterate. 

\footnotesize
\begin{center}
\begin{tabular}[h]{l}\
\ \\
\textbf{Algorithm 2.} SCP-Based Rate Control for SVC Streams\\
\hline
\ \\
\textbf{A. Initialization:}\\
\quad \textbf{A.1.} Choose an arbitrary feasible vector $\boldsymbol{\tilde x}^{(0)}\in\boldsymbol{\tilde{\mathcal D}}$.\\
\quad \textbf{A.2.} Establish problem $\textsf{P}(\boldsymbol{\tilde x}^{(0)})$.\\
\quad \textbf{A.3.} Initialize $\texttt{th}_1$ and $\texttt{th}_2$.\\
\quad \textbf{A.4.} Set $k=0$.\\
\ \\
\textbf{B. Main Loop (Outer Iteration):}\\
Until $\max_s |x_s^{(k+1)}-x_s^{(k)}|\leq \texttt{th}_1$, at the $k$th outer iterate do\\
\quad \textbf{B.1.} Establish problem  $\textsf{P}(\boldsymbol{\tilde x}^{(k)})$.\\
\quad \textbf{B.2.} Initialize $\gamma_k$ and $\boldsymbol\mu^{(0)}$.\\ 
\quad \textbf{B.3.} Set $\boldsymbol{v}^{(0)}=\boldsymbol{\tilde x}^{(k)}$.\\
\quad \textbf{B.4.} Set $t=0$.\\
\quad \textbf{B.5. Inner Iteration:} \\
\quad Until $\max_s |v_s^{(t+1)}-v_s^{(t)}|\leq \texttt{th}_2$, at the $t$th inner iterate do\\
\quad\quad \textbf{B.5.1.} For each link $l$, update $\mu_l^{(t)}$  using (\ref{eq:dual-update}).\\
\quad\quad \textbf{B.5.2.} For each source $s$, obtain $\mu^{s(t)}=\sum_{l} R_{ls}\mu_l^{(t)}$.\\
\quad\quad \textbf{B.5.3.} For each source $s$, find $i^{(t+1)}_s$ such that \\
\quad\quad\qquad $\tilde x^{(k)}_sL_{si^{(t+1)}_s}\leq\mu^{s(t)}\leq \tilde x^{(k)}_s U_{si^{(t+1)}_s}$,\\
\quad\qquad where $L_{si^{(t+1)}_s}$ and $U_{si^{(t+1)}_s}$ are calculated similar to (\ref{eq:coefficient_4})-(\ref{eq:coefficient_5}). \\
\quad\quad \textbf{B.5.4.} For each source $s$, calculate $v^{(t+1)}_s$ using (\ref{eq:vs_iteration_t}).\\
\quad\quad \textbf{B.5.5.} Let $\boldsymbol v^*=\boldsymbol v^{(t+1)}$.\\
\quad\quad \textbf{B.5.6.} Set $t\leftarrow t+1$ and repeat.\\
\quad \textbf{B.6.} Let $\boldsymbol{\tilde x}^{(k+1)}=\boldsymbol{v}^{*}$.\\
\quad \textbf{B.7.} Calculate $\boldsymbol{x}^{(k+1)}$ using (\ref{eq:xs_opt}).\\
\quad \textbf{B.8.} Set $k\leftarrow k+1$ and repeat.\\
\\
\hline
\ \\
\end{tabular}
\end{center}
\normalsize

The algorithm is initialized with a starting feasible vector $\boldsymbol{\tilde x}^{(0)}$ as well as $\texttt{th}_1$ and $\texttt{th}_2$ to check the stopping conditions for outer and inner iterations, respectively\footnote{Since both SCP procedure (Algorithm 1) and gradient projection algorithm are not finitely-convergent, one has to set a stopping criterion.}. Then, it proceeds as follows. At each outer iterate $k$, we establish problem $\textsf{P}(\boldsymbol{\tilde x}^{(k)})$ defined in (\ref{eq:opt_transformed_affine_constr})-(\ref{eq:const_affined}).  
The output of the $k$th outer iterate is the unique maximizer of problem $\textsf{P}(\boldsymbol{\tilde x}^{(k)})$, which is denoted by $\boldsymbol{\tilde x}^{(k+1)}$. At the $k$th outer iterate, problem $\textsf{P}(\boldsymbol{\tilde x}^{(k)})$ will be solved through its dual $\textsf{D}_k$. As in previous subsections we employed gradient projection algorithm to iteratively solve $\textsf{D}_k$, inside the $k$th outer iterate, we will have an iterative procedure (inner iteration). In this respect, in the $t$th inner iterate, we will update dual variable using (\ref{eq:dual-update}).  

\subsection{Algorithm 3: Simplified Algorithm}
In order to tailor the two-tier algorithm proposed in the previous subsection as Algorithm 2, here we present a simplified and efficient variant of that. Such a simplified variant relies on just the outer iteration: Indeed, instead of solving the problem $\textsf{P}(\boldsymbol{\tilde x}^{(k)})$ in the $k$th outer iterate through many iterations to obtain the optimal dual variable vector accurately, we just solve it with one iteration to obtain an approximate value of optimal dual variable vector. This is similar to choosing a quite large value of $\texttt{th}_2$ in Algorithm 2, such that after just one iteration, the stopping criterion will be satisfied. 

Note that at the $k$th outer iterate, we solve $\textsf{P}(\boldsymbol{\tilde x}^{(k)})$ using dual method to obtain $\boldsymbol{\tilde x}^{(k+1)}$. Then, this will be employed to establish $\textsf{P}(\boldsymbol{\tilde x}^{(k+1)})$. Now let $\boldsymbol{\hat x}^{(k+1)}$ be an approximate to $\boldsymbol{\tilde x}^{(k+1)}$. Then, one would come up with problem $\textsf P(\boldsymbol{\hat x}^{(k+1)})$ instead of $\textsf P(\boldsymbol{\tilde x}^{(k+1)})$. As both the aforementioned problems strive to approximate the original D.C. problem, in the worst case, $\textsf P(\boldsymbol{\hat x}^{(k+1)})$ will yield a worse approximation than $\textsf P(\boldsymbol{\tilde x}^{(k+1)})$, and the convergence guarantee will not be violated at all. The corresponding algorithm is listed as Algorithm 3. Although the convergence speed of Algorithm 3 might be slower than that of Algorithm 2, the former lends itself better for distributed implementation as it only has the outer iteration.

\begin{figure*}
\begin{eqnarray}
\label{eq:vs_iteration_t}
v_s^{(t+1)}=\frac{A_{si^{(t+1)}_{s}}}{2u_{s(i^{(t+1)}_{s}+1)}}
\left(\Delta u_{s(i^{(t+1)}_s+1)}\sqrt{1+\frac{4w_s\alpha_s \tilde x_s^{(k)} u_{s(i^{(t+1)}_s+1)}}{\Delta u_{s(i^{(t+1)}_s+1)}\mu^{s(t)}A_{si^{(t+1)}_s}}}-u_{si^{(t+1)}_{s}}-u_{s(i^{(t+1)}_{s}+1)}\right)
\end{eqnarray}
\hrule
\hrulefill
\vspace*{4pt}
\end{figure*}

%

\subsection{Message Passing Mechanisms}
The developed algorithms thus far mentioned require some form of message passing at each iterate. The first message passing is required for communicating the updated dual variables to the corresponding sources, and the second one is required for communicating the calculated source rates to the links on the corresponding paths. 
The first message passing is to be implemented by explicit message passing inspired by the work \cite{Balancing_chiang}, with messages containing updated dual variables for source rate calculations in the next iterate. 

In most previous studies, the second message passing have been accomplished implicitly by measuring the flow of link $l$. 
In our case, however, dual variable update is done using transformed domain rates $\tilde x_s,s\in\mathcal S$, and as dual variable update points out, the update cannot be accomplished by measuring the flow of links. Therefore, explicit message passing is required. 


\footnotesize
\begin{center}
\begin{tabular}[h]{l}\
\ \\
\textbf{Algorithm 3.} Distributed Rate Control for SVC Streams\\
\hline
\ \\
\textbf{A. Initialization:}\\
\quad \textbf{A.1.} Choose two arbitrary feasible vectors $\boldsymbol{\tilde x}^{(0)},\boldsymbol{\tilde x}^{(1)}\in\boldsymbol{\tilde{\mathcal D}}$.\\
\quad \textbf{A.2.} Choose a Lagrange multiplier vector $\boldsymbol{\mu}^{(1)}$.\\
\quad \textbf{A.3.} Initialize $\gamma$ and $\texttt{th}$.\\
\quad \textbf{A.4.} Set $k=1$.\\
\ \\
\textbf{B. Main Loop:}\\
Until $\max_s |x_s^{(k+1)}-x_s^{(k)}|\leq \texttt{th}$, at the $k$th iterate do\\
\quad \textbf{B.1.} For each link $l$, update $\mu_l^{(k)}$ using\\
\quad $\mu_l^{(k+1)}=$\\
\qquad $\bigg[\mu_l^{(k)}-\gamma\left\{c_l-\sum_s \frac{R_{ls}}{\alpha_s}\left(\log \tilde x_s^{(k-1)}+\frac{\tilde x_s^{(k)}-\tilde x^{(k-1)}_s}{\tilde x^{(k-1)}_s}\right)\right\}\bigg]^+$.\\
\quad \textbf{B.2.} Obtain $\mu^{s(k)}=\sum_{l} R_{ls}\mu_l^{(k)}$.\\
\quad \textbf{B.3.} For each source $s$, find $i^{(k+1)}_s$ such that \\
\quad\quad\qquad $\tilde x^{(k)}_sL_{si^{(k+1)}_s}\leq\mu^{s(k)}\leq \tilde x^{(k)}_s U_{si^{(k+1)}_s}$,\\
\quad\quad  where $L_{si^{(k+1)}_s}$ and $U_{si^{(k+1)}_s}$ are calculated by (\ref{eq:coefficient_4})-(\ref{eq:coefficient_5}). \\
\quad \textbf{B.4.} For each source $s$, calculate $x^{(k+1)}_s$ using (\ref{eq:tildexs_opt}) and (\ref{eq:xs_opt}).\\
\quad \textbf{B.5.} Set $k\leftarrow k+1$ and repeat.\\
\\
\hline
\ \\
\end{tabular}
\end{center}
\normalsize

%

\subsection{Convergence}
We conclude this section by a result on the convergence of the proposed algorithms.
\begin{myTheo}
If step sizes in Algorithm 2 are chosen sufficiently small, then starting from any feasible vector $\boldsymbol{\tilde x}^{(0)}$, Algorithm 2 converges to locally optimal points of problem (\ref{eq:opt_transformed}).
\end{myTheo}
\begin{proof}
(\emph{Sketch}) The proof of this theorem is quite similar to the proof of Theorem 1 of \cite{Evans_TSP}, and we omit the details here due to space limit. 

To prove the assertions of theorem, similar to \cite{Evans_TSP}, one must show that the sequence $\{\boldsymbol{\tilde x}^{(k)}\}$ is feasible and also converges to a KKT point. It can be shown that if $\boldsymbol{\tilde x}^{(0)}$ is chosen to be feasible, provided that step sizes are chosen to be sufficiently small, then $\boldsymbol{\tilde x}^{(k)},\forall k$ will remain feasible. The next step is to show that $\{\boldsymbol{\tilde x}^{(k)}\}$ is a nondecreasing upper-bounded sequence and thereby it will converge asymptotically. 

%
\end{proof}

\textit{Remark:} Note that $\texttt{th}_2$ has no effect on the convergence guarantee of Algorithm 2 as for sufficiently small step size, all primal values within the inner iteration remain feasible. Therefore, the above result will be applicable to Algorithm 3, as well.

\section{Simulation Experiments}
\label{sec:simulation}
In this section, we evaluate the performance of Algorithm 3 for SVC streams through extensive simulation experiments. First, we introduce SVC transmission scenarios that will be used in our experiments. It is followed by three simulation scenarios.  

\subsection{Simulation Setup for SVC} 
\label{sec:mmscenario}
In order to achieve realistic simulation results, we adopt testing conditions for SVC standard provided in \cite{SVCTest}. We have used four different video sequences: \emph{Football}, \emph{Foreman}, \emph{Bus}, and \emph{Mobile} \cite{SVC-sequence} to take into account the requirements of a broad range of video applications.
These sequences can be transmitted in different settings, each one having specific features such as number of spatial, temporal, and quality layers. Table \ref{table_vs} summarizes the required bit rate for transmission of each video sequence with different settings in terms of the number of spatial, temporal, and quality layers.

\begin{table}[h]
\caption{Tested Bit Rates for the Quality and Spatial Scalability Test}
\label{table_vs}
\centering 
\begin{tabular}{|c c rrr |} 
\hline 
\multicolumn{1}{|c} \textbf{Sequence} &\textbf{Format} &\multicolumn{3}{c|}{\textbf{Bit Rates (Kbps)}}
\\ [0.5ex]
\hline 
& \textbf{QCIF} 15 Hz   &96 & 128 & 192  \\[-1ex]
\raisebox{1.5ex}{\emph{Bus}} & 
\textbf{CIF} 30 Hz &  384 & 512 & 768  \\[1ex]
& \textbf{QCIF} 15 Hz    &192 & 256 & 384  \\[-1ex]
\raisebox{1.5ex}{\emph{Football}} & 
\textbf{CIF} 30 Hz &  768 & 1024 & 1536  \\[1ex]
& \textbf{QCIF} 15 Hz   &48 & 64 & 96  \\[-1ex]
\raisebox{1.5ex}{\emph{Foreman}} & 
\textbf{CIF} 30 Hz &  192 & 256 & 384  \\[1ex]
& \textbf{QCIF} 15 Hz   &64 & 96 & 128  \\[-1ex]
\raisebox{1ex}{\emph{Mobile}} & 
\textbf{CIF} 30 Hz &  256 & 384 & 512  \\[1ex]
\hline 
\end{tabular}
\end{table}

\subsection{Scenario 1}
For the first scenario, we consider a simple scenario, in which 12 SVC-encoded video streams share a bottleneck link with capacity $c=5$ Mbps. Video Sequences and weights of all sessions are listed in Table \ref{table_f1}. All sources have the same parameter $\alpha=2$. 
For the sake of illustration, we let all sources have utility functions with the same quality index sequence $\{u_{si}\},s=1,\dots,12$. However, their rate index sequences $\{\beta_{si}\},s=1,\dots,12$ are determined based on the video sequence (Table \ref{table_vs}). 
We choose the following quality index sequence
$$\{u_{si}\}=\{0,2,2.8,3.4,3.9,4.3\};\quad s=1,\dots,12$$ 
It's easy to verify that $\{u_{si}\}_{i=0,\dots,5}$ is a monotonically increasing and strictly concave sequence and $G_{si}\gg 1$ and therefore the conditions of Lemma \ref{concave_Lemma} are satisfied. 

\begin{figure}[t]
\centering
\subfigure[Rate Allocation for SVC Streams]{
\includegraphics[scale=0.4]{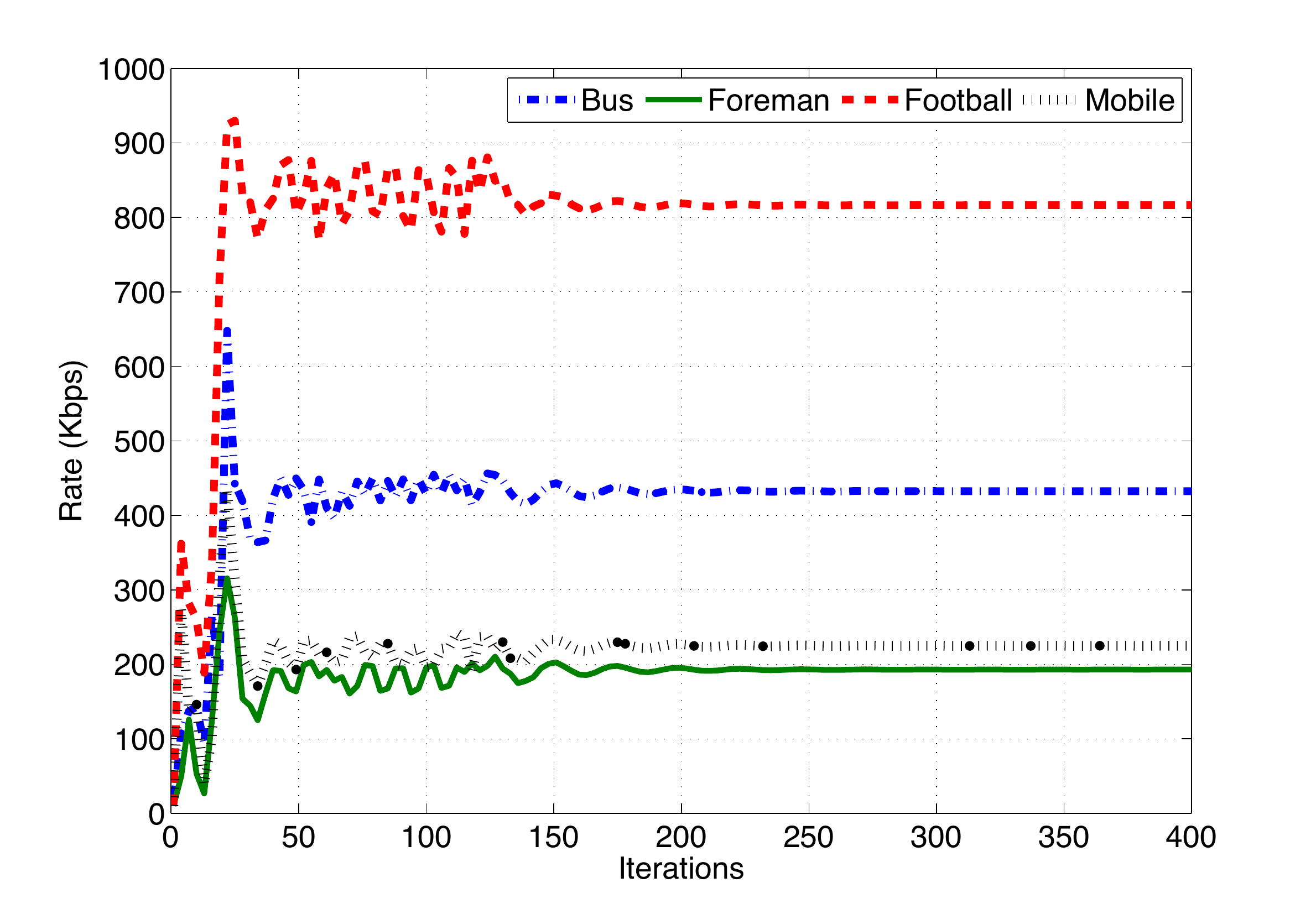}
\label{fig_rate_f1}}
\subfigure[Evolution of Dual Variable]{
\includegraphics[scale=0.4]{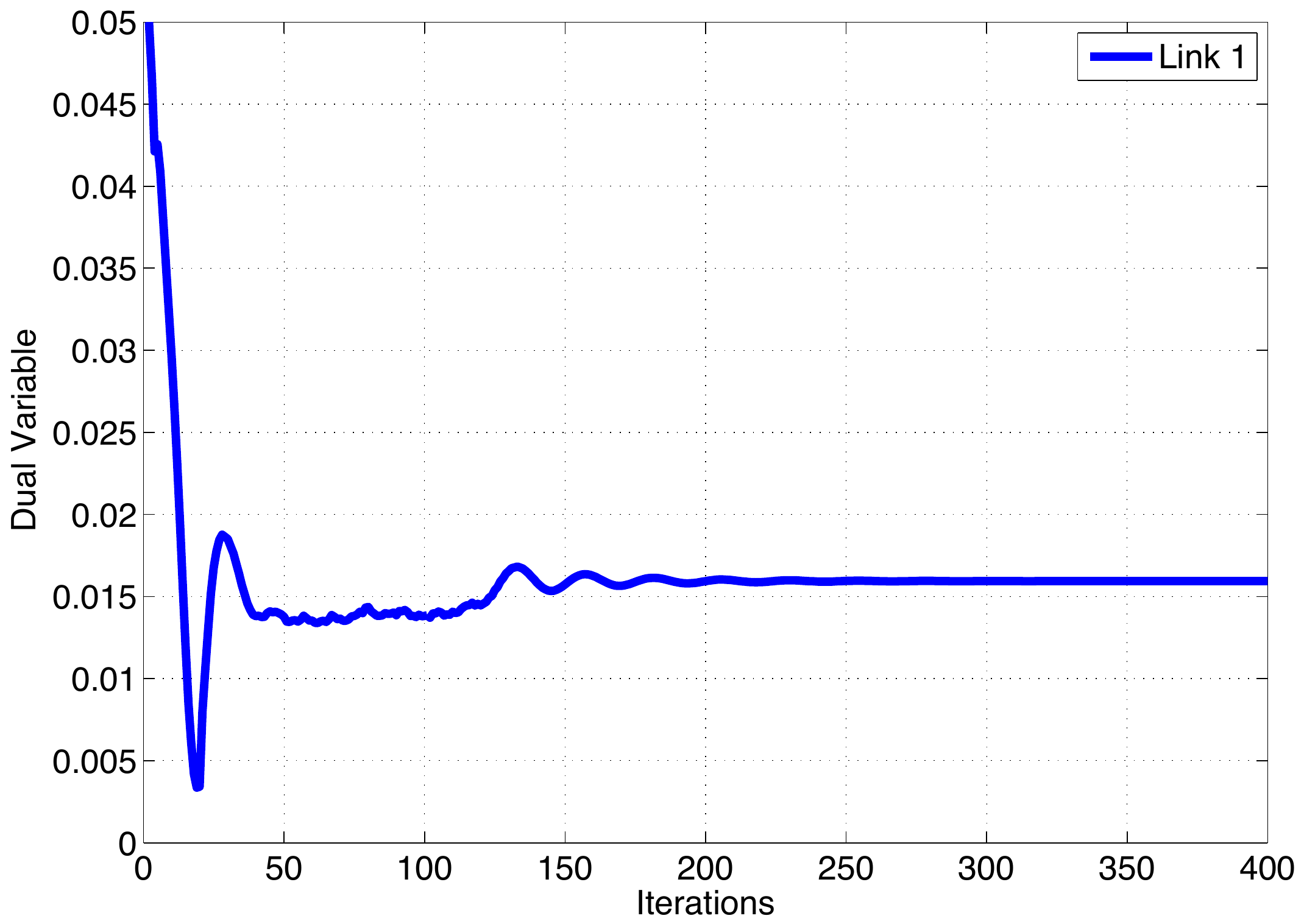}
\label{fig_p_f1}}

\label{fig_top_conf}
\caption{Simulation Results for Scenario 1}
\end{figure}

The rate allocation is carried out using Algorithm 3 with step size parameter $\gamma=10^{-2}$. The allocated rates obtained from Algorithm 3 for  video sessions are summarized in Table \ref{table_f1}. Fig. \ref{fig_rate_f1} and \ref{fig_p_f1} depict the evolution of source rates and the dual variable, respectively, as the transient behavior of the algorithm. Assessment of such a transient behavior gives insights into how much fast the algorithm converges toward global optimality. Both figures reveal that the convergence is quite fast and about 200 iteration steps are needed to reach the steady state. 

\begin{table}[b]
\caption{Rate Allocation for Scenario 1}
\label{table_f1}
\begin{center}
\begin{tabular}{|c|c|c|c|c|}
\hline
\multicolumn{1}{|c|}{ \textbf{Source} } & \multicolumn{1}{c|}{\textbf{Video Sequence}} &
\multicolumn{1}{c|}{ $\boldsymbol{w_s}$ } & \multicolumn{1}{c|}{ $\boldsymbol{x_s}$ \textbf{(Kbps)} } &
\multicolumn{1}{c|}{ $\boldsymbol{i^*_s}$ } \\\hline
$\boldsymbol{1,2,3}$  &  \emph{Bus}  &  $2$    &  $432$ &  $4$\\\hline
$\boldsymbol{4,5,6}$  &  \emph{Foreman}  &  $0.75$   &  $192$ &  $3$\\\hline
$\boldsymbol{7,8,9}$  &  \emph{Football}  &  $2$    &  $816$ &  $4$\\\hline
$\boldsymbol{10,11,12}$  &  \emph{Mobile}  &  $1.5$    &  $224$ &  $3$\\\hline
\end{tabular}
\end{center}
\end{table}

\subsection{Scenario 2: The Effect of Quality Indices}
\begin{figure}[t]
\centering
\subfigure[Utility Function of SVC Streams]{
\includegraphics[scale=.4]{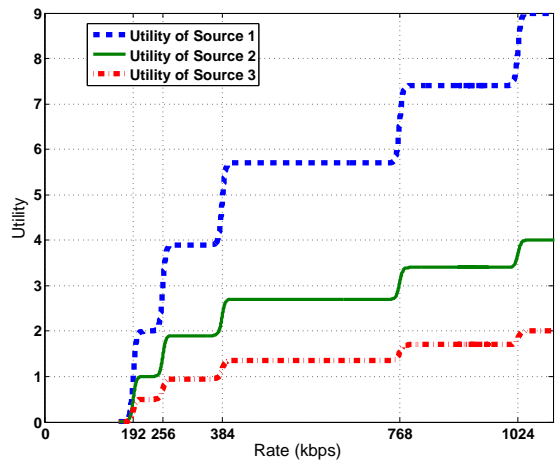}
\label{fig_um_s2}}
\subfigure[Rate Allocation for SVC Streams]{
\includegraphics[scale=0.4]{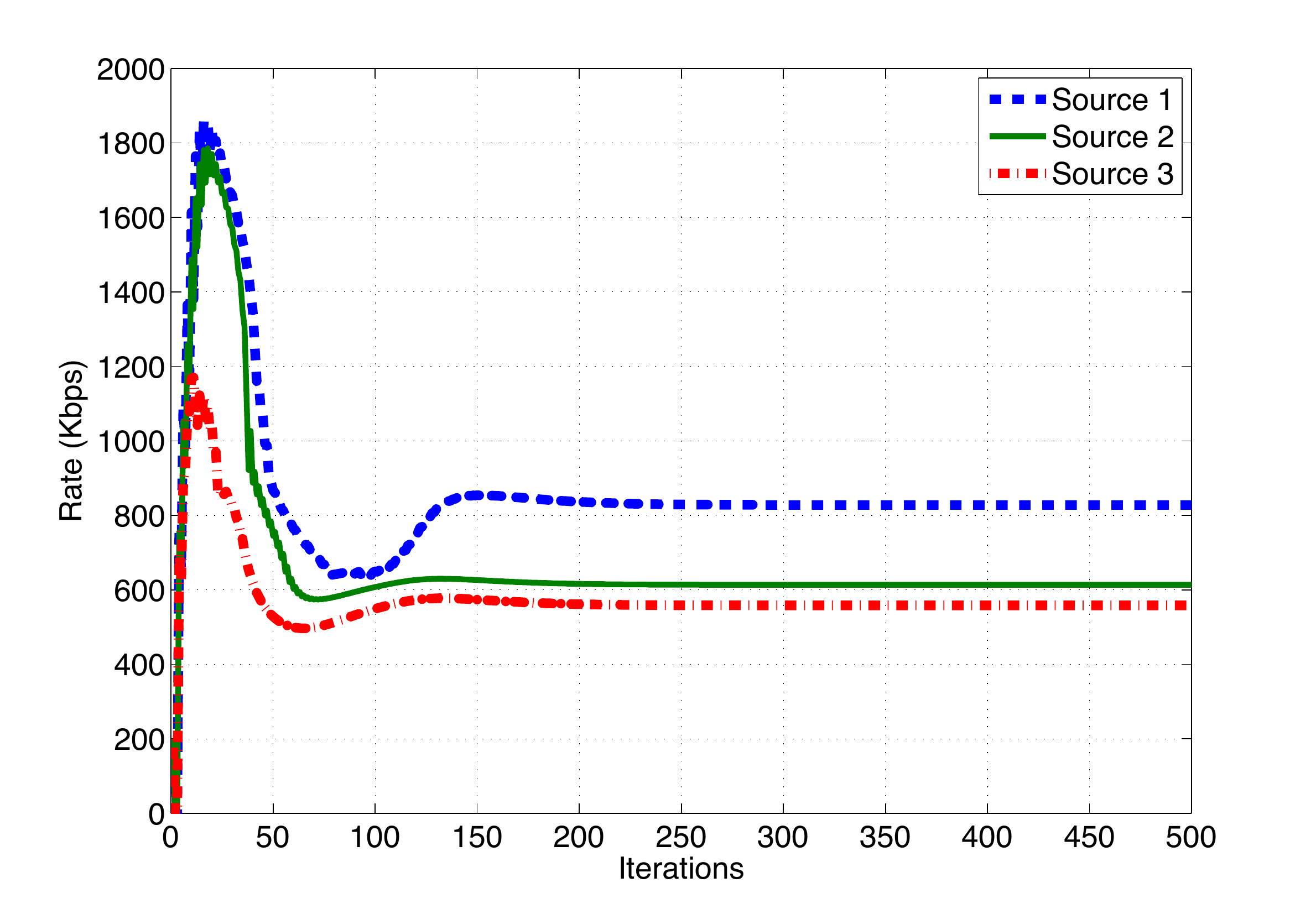}
\label{fig_rate_s2}}
\subfigure[Evolution of Dual Variable]{
\includegraphics[scale=0.4]{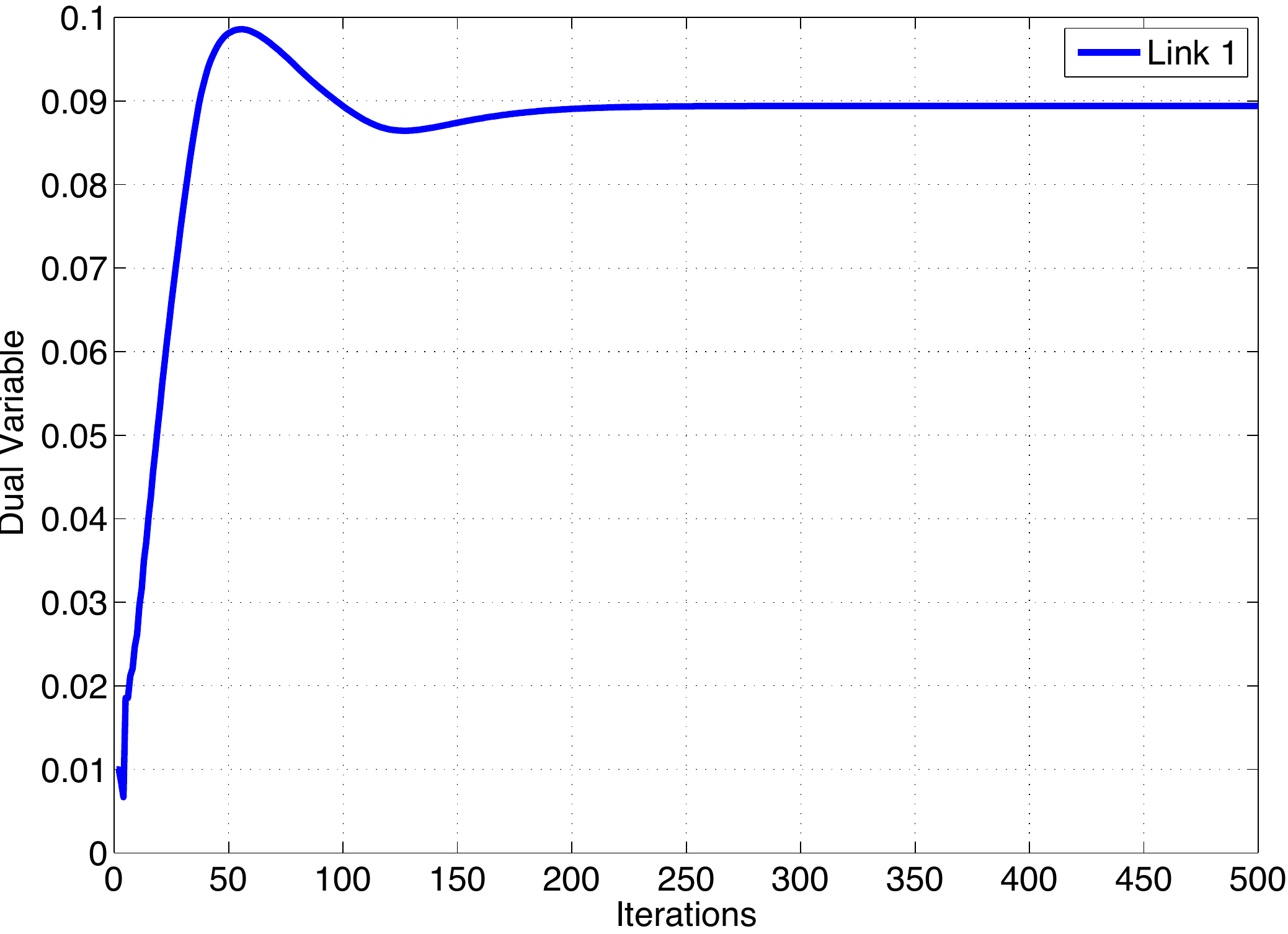}
\label{fig_p_s2}}

\label{fig_top_conf}
\caption{Simulation Results for Scenario 2}
\end{figure}

In this scenario we study the impact of different quality indices on rate allocation to account for existence of different preferences amongst users. Towards this, we consider a topology in which 3 sources share a bottleneck link with capacity $c=2$ Mbps. Then, we associate \textit{Football} sequence with the same weight parameters to all sources, with different quality index sequences listed as follows
$$\{u_{1i}\}=\{0,2,3.9,5.7,7.4,9,10.5\}$$ 
$$\{u_{2i}\}=\{0,1,1.9,2.7,3.5,4.2,4.8\}$$ 
$$\{u_{3i}\}=\{0,0.5,0.95,1.35,1.7,1.95,2.15\}$$ 
It is straightforward to show these sequences are monotonically increasing and strictly concave and $G_{si}\gg 1$.
For the sake of illustration, the multimodal sigmoid approximated utility functions corresponding to these video sessions with abovementioned quality indices are depicted in Fig. \ref{fig_um_s2}. The value of $\alpha$ for all sources is set to 3 and step size is chosen to be $\gamma=10^{-2}$. 
The allocated rates obtained from Algorithm 3 for video sessions are summarized in Table \ref{table_s2}.

\begin{table}[b]
\caption{Rate Allocation for Scenario 2}
\label{table_s2}
\begin{center}
\begin{tabular}{|c|c|c|c|c|}
\hline
\multicolumn{1}{|c|}{ \textbf{Source} } & \multicolumn{1}{c|}{\textbf{Video Sequence}} &
\multicolumn{1}{c|}{ $\boldsymbol{w_s}$ } & \multicolumn{1}{c|}{ $\boldsymbol{x_s}$ \textbf{(Kbps)} } &
\multicolumn{1}{c|}{ $\boldsymbol{i^*_s}$ } \\\hline
$\boldsymbol{1}$  &  \emph{Football}  &  $1$    &  $828$ &  $4$\\\hline
$\boldsymbol{2}$  &  \emph{Football}  &  $1$   &  $614$ &  $3$\\\hline
$\boldsymbol{3}$  &  \emph{Football}  &  $1$    &  $558$ &  $3$\\\hline
\end{tabular}
\end{center}
\end{table}
Fig. \ref{fig_rate_s2} and \ref{fig_p_s2} display the evolution of source rates and dual variable, respectively. Focusing on Fig. \ref{fig_rate_s2}, it is clear that the first video source whose utility function has the most rapidly increasing envelope, achieves more rate than other sources. This fact is consistent with NUM formulation, where the optimization problem allocates available bandwidth so that utility functions with higher increases attain larger rates in order to maximize the total utility as much as possible. However, this is in contrast to rate allocation for SVC streams with utility-proportional rate allocation \cite{Talebi-Elsevier}, in which rate allocation is carried out in favor of sources with utility functions having slowly increasing envelope.

\subsection{Scenario 3: The Multiple Link Topology}
\begin{figure}[t]
\centering
\subfigure[Topology of Scenario 3]{
\includegraphics[scale=.4]{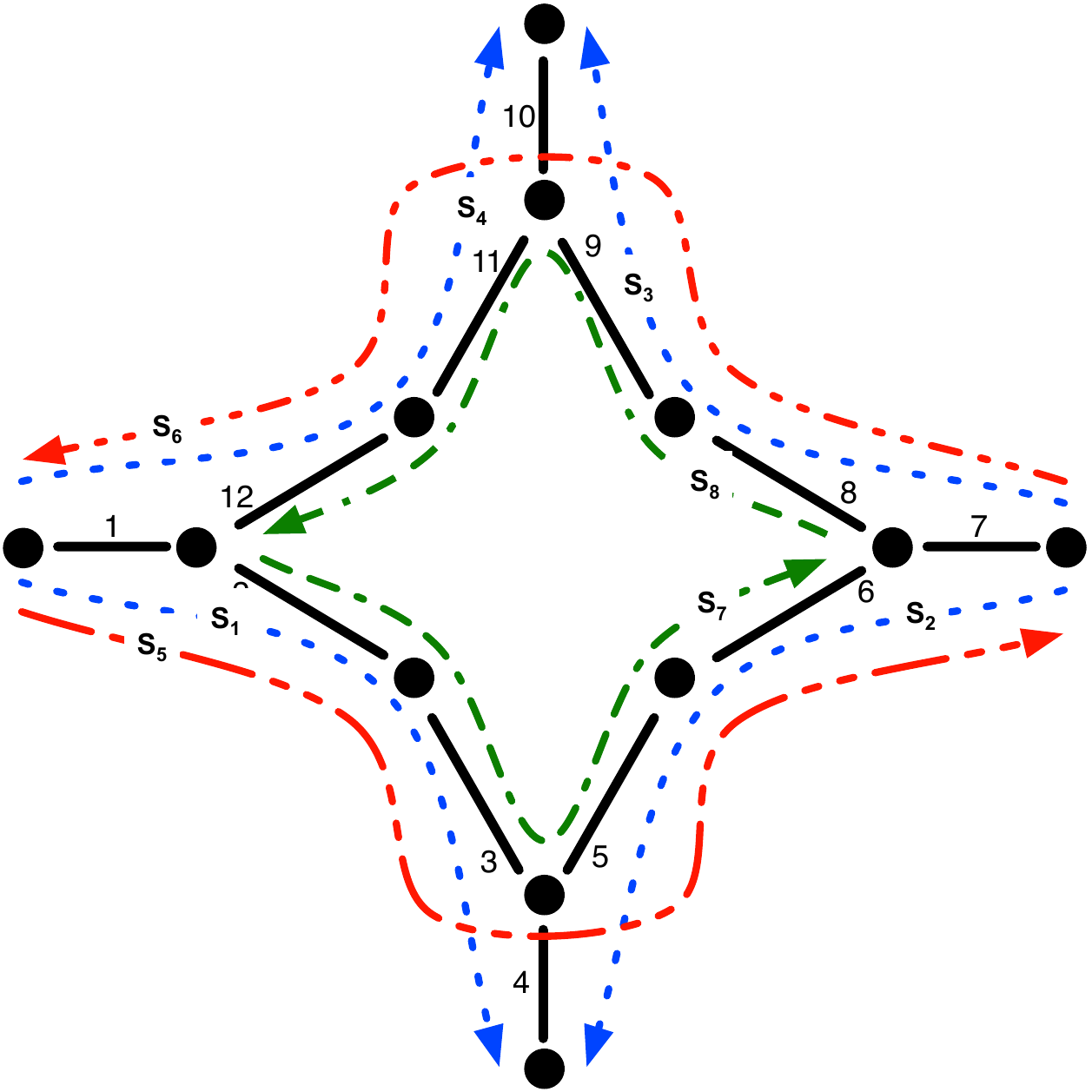}
\label{fig_topology}}
\subfigure[Rate Allocation for SVC Streams]{
\includegraphics[scale=0.4]{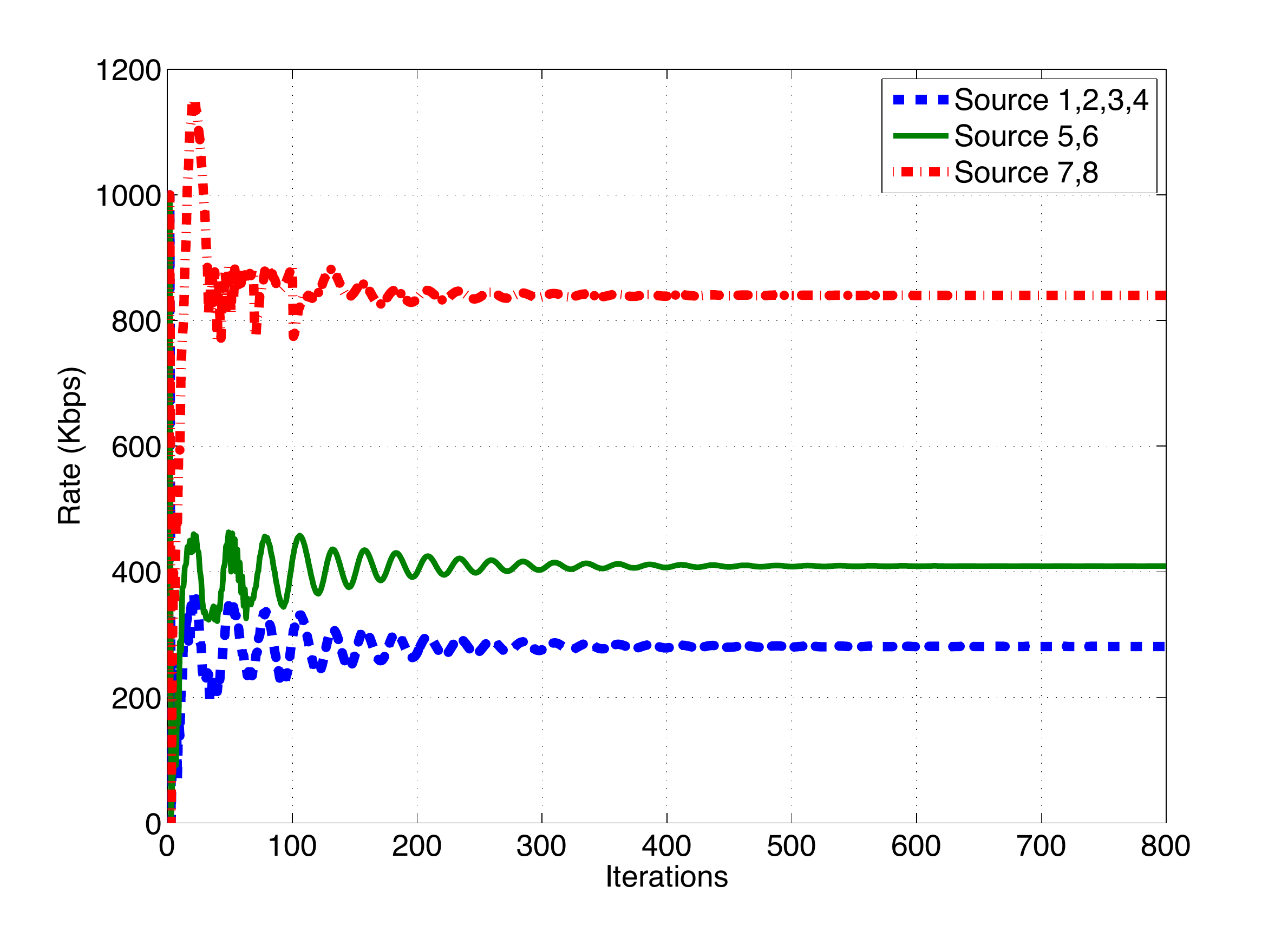}
\label{fig_rate_s3}}
\subfigure[Evolution of Dual Variables]{
\includegraphics[scale=0.4]{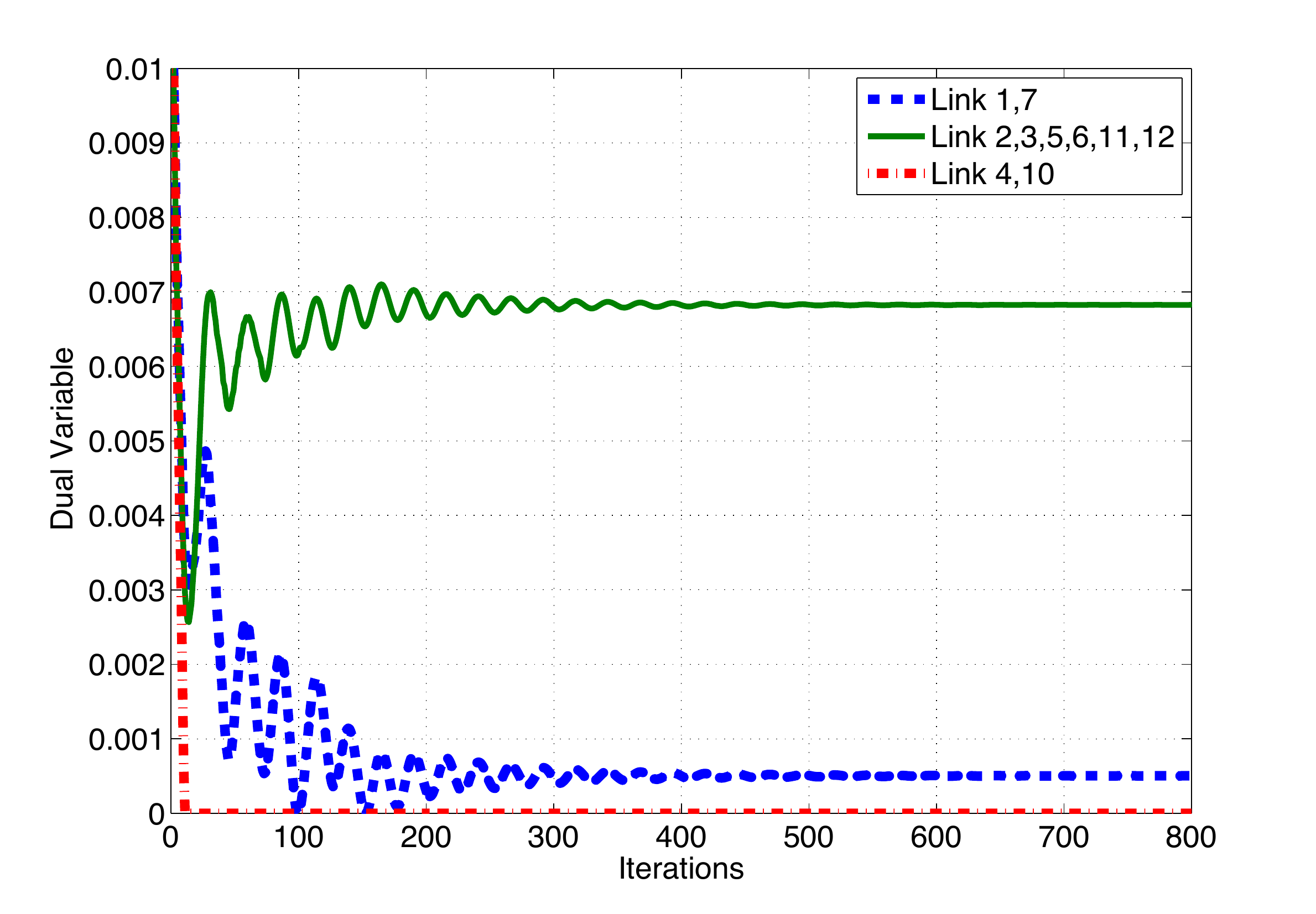}
\label{fig_p_s3}}

\label{fig_top_conf}
\caption{Simulation Results for Scenario 3}
\end{figure}

For the third scenario, we focus on a network whose topology has several bottleneck links. The network has 8 video sessions traversing 12 links and whose topology along with paths of video sessions is depicted in Fig. \ref{fig_topology}, in which for the sake of brevity, only links with more than one video session are depicted. The capacity of all links is set to $1$ Mbps. For all video sessions, we choose $\alpha=5$ and the quality index sequence as follows
$$\{u_{si}\}=\{0, 2, 2.8, 3.4, 3.9, 4.3\}, \qquad s=1,\dots,8$$ 

\begin{table}[t]
\caption{Rate Allocation for Scenario 3}
\begin{center}
\begin{tabular}{|c|c|c|c|c|c|}
\hline
\multicolumn{1}{|c|}{ \textbf{Source} } & \multicolumn{1}{c|}{\textbf{Video Seq.}} &
\multicolumn{1}{c|}{ $\boldsymbol{w_s}$ } & \multicolumn{1}{c|}{ $\boldsymbol{x_s}$ \textbf{(Kbps)} }
& \multicolumn{1}{c|}{ $\boldsymbol{i^*_s}$ }
 \\\hline

 $\boldsymbol{1,2,3,4}$  &  \emph{Mobile}  &  $1$  &   $281$ & 4\\\hline
 $\boldsymbol{5,6}$  &  \emph{Bus}  &  $2$  &  $409$ & 4\\\hline
 $\boldsymbol{7,8}$  &  \emph{Football}  &  $2$  &  $840$ &4\\\hline
\end{tabular}
\end{center}
\label{table_s3}
\end{table}

The results of rate allocation for this scenario with $\gamma= 10^{-2}$ is summarized in Table \ref{table_s3}. In order to investigate the convergence behavior of this scenario, the evolution of session rates is depicted in Fig. \ref{fig_rate_s3}. This figure demonstrates that similar to Scenario 1 and Scenario 2 (with single bottleneck link), convergence is relatively fast. 
The evolution of dual variables is shown in Fig. \ref{fig_p_s3}. As this figure presents, some links have non-zero dual variables that implies the saturation of such links. 

\section{Conclusion}
\label{sec:conclusion}
In this paper we addressed rate allocation for scalable video streaming applications via Network Utility Maximization (NUM) framework. 
To take the advantage of current developments in NUM problems, we first remedied nonsmoothness of the NUM for such applications by exploiting arbitrarily tight sigmoid-based approximations of the ideal discretely adaptive utility functions. We then presented a nonconvex D.C. problem whose solution is shown to be the solution of an approximation of the smoothed NUM. 
The presented D.C. problem comprised strictly concave objective and D.C. constraints. We established the concavity of the objective of such a D.C. problem based on some mild conditions on the quality indices as the utility function parameters. 
As such, by employing Sequential Convex Programming (SCP) approach, we replaced the problem with a sequence of strictly convex programs. 
Taking the benefit of SCP approach for the noncoonvex D.C. problem, we devised two iterative distributed algorithms, 
for which convergence to a KKT point of the nonconvex D.C. problem is guaranteed under mild conditions. 
The algorithms were based on solving the dual of convex programs generated by SCP approach using gradient projection algorithm. As such, the two algorithms lend themselves to distributed implementation with low message passing overhead. 
Our experimental results have shown that the proposed algorithms have tractable convergence properties.
A promising future direction to our development is to consider jointly optimal rate control and scheduling for scalable video streaming over multihop wireless networks.

\begin{figure*}
\begin{eqnarray}
&&\lim_{\epsilon_1\rightarrow 0} \tilde U^\prime(z)>\lim_{\epsilon_2\rightarrow 0} \tilde U^\prime(z)\nonumber\\
&\Longleftrightarrow&\frac{A_{i}-B_{i}}{(A_iG_{i+1}+A_{i})(A_iG_{i+1}+B_{i})}> \frac{A_{i+1}-B_{i+1}}{(A_iG_{i+1}+A_{i+1})(A_iG_{i+1}+B_{i+1})}\nonumber\\
\label{eq:proof_tmp}
&\Longleftrightarrow&\frac{A_{i}\left(1-\frac{u_{i}}{u_{i+1}}\right)}{A_i^2\left(G_{i+1}+1\right)\left(G_{i+1}+\frac{u_{i}}{u_{i+1}}\right)}> \frac{A_{i+1}\left(1-\frac{u_{i+1}}{u_{i+2}}\right)}{\left(A_iG_{i+1}+A_{i+1}\right)\left(A_iG_{i+1}+\frac{u_{i+1}}{u_{i+2}}A_{i+1}\right)}\nonumber\\
\end{eqnarray}
\hrule
\hrulefill
\begin{eqnarray}
&&\frac{A_{i}\left(1-\frac{u_{i}}{u_{i+1}}\right)}{A_i^2\left(G_{i+1}+1\right)\left(G_{i+1}+\frac{u_{i}}{u_{i+1}}\right)}>
\frac{A_{i}G_{i+1}^2\left(1-\frac{u_{i+1}}{u_{i+2}}\right)}{\left(A_iG_{i+1}+A_{i}G_{i+1}^2\right)\left(A_iG_{i+1}+\frac{u_{i+1}}{u_{i+2}}A_{i}G_{i+1}^2\right)}\nonumber\\
&\Longleftrightarrow&\frac{1-\frac{u_{i}}{u_{i+1}}}{G_{i+1}+\frac{u_{i}}{u_{i+1}}}>
\frac{1-\frac{u_{i+1}}{u_{i+2}}}{1+G_{i+1}\frac{u_{i+1}}{u_{i+2}}}\nonumber\\
&\Longleftrightarrow&\frac{u_{i+1}-u_{i}}{G_{i+1}u_{i+1}+u_{i}}>
\frac{u_{i+2}-u_{i+1}}{u_{i+2}+G_{i+1}u_{i+1}}\nonumber\\
&\Longleftrightarrow&\left(u_{i+1}-u_{i}\right)\left(u_{i+2}+G_{i+1}u_{i+1}\right)-\left(u_{i+2}-u_{i+1}\right)\left(G_{i+1}u_{i+1}+u_{i}\right)> 0\nonumber\\
\label{eq:proof_tmp2}
&\Longleftrightarrow&u_{i+1}\left(1-G_{i+1}\right)\left(u_{i+2}+u_{i}\right)+2G_{i+1}u_{i+1}^2-2u_{i+2}u_i> 0
\end{eqnarray}
\hrule
\hrulefill
\end{figure*}

\section*{Appendix I: Proof of Lemma 1}
\label{sec:concave_Lemma_proof}
Since $\tilde U$ is not guaranteed to be twice differentiable, we cannot use its second derivative to establish concavity conditions. However, as $G_i\gg 1,i=1,\dots,N$ results in differentiability of $\tilde U$, we use the first order strict concavity criterion to derive the conditions under which $\tilde U$ is strictly concave. In other words, we seek conditions under which $\tilde U^\prime(z)$ is monotonically decreasing. 
Considering (\ref{eq:transformed_utility_explicit}), $\tilde U^\prime$ for $z\in\mathbf{int}\textrm{ }\mathcal{\tilde I}_i$ is given by
\begin{eqnarray}
&&\frac{d}{dz}\log\left(\frac{\Delta u_{i+1}z}{z+e^{\alpha\beta_{i}}}+u_{i}\right)\nonumber\\
&&\quad=\frac{\Delta u_{i+1}e^{\alpha\beta_{i}}}{\left(z+e^{\alpha\beta_{i}}\right)^2\left(\frac{\Delta u_{i+1}z}{z+e^{\alpha\beta_{i}}}+u_{i}\right)}\nonumber\\
&&\quad=\frac{(u_{i+1}-u_i)e^{\alpha\beta_{i}}}{\left(z+e^{\alpha\beta_{i}}\right)\left(\Delta u_{i+1}z+u_iz+u_ie^{\alpha\beta_{i}}\right)}\nonumber\\
&&\quad=\frac{(u_{i+1}-u_i)A_i}{\left(z+A_i\right)\left(u_{i+1}z+u_iA_i\right)}\nonumber\\
&&\quad=\frac{\frac{u_{i+1}-u_i}{u_{i+1}}A_i}{\left(z+A_i\right)\left(z+\frac{u_i}{u_{i+1}}A_i\right)}\nonumber\\
&&\quad=\frac{A_i-B_i}{\left(z+A_i\right)\left(z+B_i\right)}\nonumber\\
\label{eq:transformed_utility_derivative_i}
&&
\end{eqnarray}
In order to obtain decreasing monotonicity conditions, for $i=0,\dots,N-1$ we require
\begin{equation}
\label{eq:mono_dec_condition_1}
\forall z_1,z_2\in\mathbf{int}\textrm{ }\mathcal{\tilde I}_i, z_1<z_2 \Rightarrow \tilde U^{\prime}(z_1)>\tilde U^{\prime}(z_2)
\end{equation}
and
\begin{equation}
\label{eq:mono_dec_condition_2}
\forall z_1\in\mathcal{\tilde I}_i,\forall z_2\in\mathcal{\tilde I}_{j>i} \Rightarrow \tilde U^{\prime}(z_1)>\tilde U^{\prime}(z_2)
\end{equation}
First we concentrate on (\ref{eq:mono_dec_condition_1}). 
Over $\mathbf{int}\textrm{ }\mathcal{\tilde I}_i$, $\tilde U^\prime(z)$ is differentiable with the derivative given by
\begin{equation}
\tilde U^{\prime\prime}(z)=\left(B_{i}-A_{i}\right)\frac{2z+A_{i}+B_{i}}{(z+A_{i})^2(z+B_{i})^2};\quad z\in\mathbf{int}\textrm{ }\mathcal{\tilde I}_i
\end{equation}
As $\mathcal{\tilde I}_i\subset\mathbb R_+$, we have $z>0$ and hence $\tilde U^\prime$ is monotonically decreasing over $\mathbf{int}\textrm{ }\mathcal{\tilde I}_i$ if and only if 
$$B_i<A_i\Longleftrightarrow u_i<u_{i+1}$$
which is equivalent to saying that $\{u_i\}_{i=0,\dots,N}$ is a monotonically increasing sequence. Fortunately, this condition is always satisfied as a higher layer has a higher quality index.

Now we focus on (\ref{eq:mono_dec_condition_2}). Since $\tilde U^\prime$ is monotonically decreasing over  $\mathbf{int}\textrm{ }\mathcal{\tilde I}_i$, it suffices to satisfy (\ref{eq:mono_dec_condition_2}) for $j=i+1$, i.e. for $z_1\in\mathcal{\tilde I}_i$ and $z_2\in\mathcal{\tilde I}_{i+1}$. We characterize $z_1$ and $z_2$ as having respectively $\epsilon_1>0$ and $\epsilon_2>0$ distance to the joint boundary of $\mathcal{\tilde I}_i$ and $\mathcal{\tilde I}_{i+1}$, i.e. $z_1=e^{\alpha\left(\beta_{i}+\Delta \beta_{i+1}/2\right)}-\epsilon_1$ and $z_2=e^{\alpha\left(\beta_{i}+\Delta \beta_{i+1}/2\right)}+\epsilon_2$. Then, (\ref{eq:mono_dec_condition_2}) must hold for some $\epsilon_1,\epsilon_2>0$. Recalling $e^{\alpha\left(\beta_{i}+\Delta \beta_{i+1}/2\right)}=A_iG_{i+1}$, calculating (\ref{eq:mono_dec_condition_2}) as $\epsilon_1$ and $\epsilon_2$ approach 0, yields (\ref{eq:proof_tmp}).
On the other hand, $$A_{i+1}=e^{\alpha\beta_{i+1}}=e^{\alpha\beta_{i}}e^{\alpha\left(\beta_{i+1}-\beta_i\right)}=A_iG_{i+1}^2$$ 
Substituting this into (\ref{eq:proof_tmp}) yields (\ref{eq:proof_tmp2}).
Then, for $G_{i+1}\gg 1$, we get
$$u_{i+1}>\frac{u_{i+2}+u_i}{2}$$
or equivalently, $u_{i+1}-u_{i}> u_{i+2}-u_{i+1}$, 
which tells us that $\{u_i\}_{i=0,\dots,N}$ is a strictly concave sequence and completes the proof.

\section*{Appendix II: Proof of Theorem 2}
Due to strict convexity of the problem, Lagrangian maximization has a unique solution, for $\boldsymbol\mu$ fixed.  
According to KKT condition, the unique maximizer to Lagrangian maximization (\ref{eq:Lagrangian}) is the stationary point of $L_k(\boldsymbol{\tilde x},\boldsymbol\mu)$, i.e. the point at which $\nabla L_k(\boldsymbol{\tilde x},\boldsymbol\mu)$ vanishes \cite{Bert_NLP}. In order to find the stationary $\boldsymbol{\tilde x}^{(k+1)}$, using (\ref{eq:transformed_utility_derivative_i}), we find the derivative of $L_k(\boldsymbol{\tilde x},\boldsymbol\lambda)$ and then solve $\nabla L_k(\boldsymbol{\tilde x}^{(k+1)},\boldsymbol\mu)=\mathbf 0$ as follows
\begin{eqnarray}
\frac{\partial L_k}{\partial \tilde x_s}&=&\frac{d}{d\tilde x_s} w_s \tilde U_s(\tilde x_s)-\sum_l\mu_l\frac{\partial \hat g_l(\boldsymbol{\tilde x},\boldsymbol{\tilde x}^{(k)})}{\partial\tilde x_s}\nonumber\\
&=& \frac{d}{d\tilde x_s} w_s \log\left(\frac{\Delta u_{s(i+1)}\tilde x_s}{\tilde x_s+e^{\alpha_s\beta_{si}}}+u_{si}\right)-\sum_{l}\frac{R_{ls}\mu_l}{\alpha_s\tilde x^{(k)}_s}\nonumber\\
&=&\frac{w_s(A_{si}-B_{si})}{\left(\tilde x_s+A_{si}\right)\left(\tilde x_s+B_{si}\right)}-\frac{\mu^s}{\alpha_s\tilde x^{(k)}_s}
\label{eq:Lagrangian_simplified}
\end{eqnarray}
where $\mu^s\triangleq\sum_l R_{ls}\mu_l$.
Optimal transformed rate of source $s$, $\tilde x^{(k+1)}_s$, is the solution of the $\frac{\partial L_k}{\partial \tilde x_s}=0$. By some algebraic manipulation on (\ref{eq:Lagrangian_simplified}), $\frac{\partial L_k}{\partial \tilde x_s}=0$ is rewritten as
\begin{eqnarray}
\label{eq:stationary_root}
\left(\tilde x^{(k+1)}_s\right)^2&+&\left(A_{si^{(k+1)}_{s}}+B_{si^{(k+1)}_{s}}\right)\tilde x^{(k+1)}_s+A_{si^{(k+1)}_{s}}B_{si^{(k+1)}_{s}}\nonumber\\
&+&\left(B_{si^{(k+1)}_{s}}-A_{si^{(k+1)}_{s}}\right)\frac{w_s\alpha_s\tilde x^{(k)}_s}{\mu^s}=0\nonumber\\
&&
\end{eqnarray}
where $i^{(k+1)}_s$ is the index of interval within which $\tilde x^{(k+1)}_s$ falls. For the sake of brevity in our derivations, we let $y=\tilde x_s^{(k+1)}$ and $j=i_s^{(k+1)}$. Then, using these new variables  (\ref{eq:stationary_root}) is rewritten as 
\begin{eqnarray}
y^2+\left(A_{sj}+B_{sj}\right)y+A_{sj}B_{sj}+\left(B_{sj}-A_{sj}\right)\frac{w_s\alpha_s\tilde x^{(k)}_s}{\mu^s}=0\nonumber
\end{eqnarray}
whose solution is given below.

\begin{figure*}
\begin{eqnarray}
y&=&\frac{-(A_{sj}+B_{sj})+\sqrt{(A_{sj}+B_{sj})^2-4A_{sj}B_{sj}-\frac{4w_s\alpha_s\tilde x^{(k)}_s}{\mu^s}(B_{sj}-A_{sj})}}{2}\nonumber\\
&=&\frac{-(A_{sj}+B_{sj})+(A_{sj}-B_{sj})\sqrt{1+\frac{4w_s\alpha_s\tilde x^{(k)}_s}{(A_{sj}-B_{sj})\mu^s}}}{2}\nonumber\\
&=&\frac{A_{sj}}{2}\left(-1-\frac{u_{sj}}{u_{s(j+1)}}+\left(1-\frac{u_{sj}}{u_{s(j+1)}}\right)
\sqrt{1+\frac{4w_s\alpha_s\tilde x^{(k)}_su_{s(j+1)}}{(u_{s(j+1)}-u_{sj})A_{sj}\mu^s}}\right)\nonumber\\
&=&\frac{A_{sj}}{2u_{s(j+1)}}\left(\Delta u_{s(j+1)}\sqrt{1+\frac{4w_s\alpha_s\tilde x^{(k)}_s u_{s(j+1)}}{\Delta u_{s(j+1)}A_{sj}\mu^s}}-u_{s(j+1)}-u_{sj}\right)\nonumber
\end{eqnarray}
Now assuming $y\in\mathcal I_{sj}$ yields
\begin{eqnarray}
&&\exp\left\{\alpha_s\left(\beta_{sj}-\frac{\Delta\beta_{sj}}{2}\right)\right\}\leq y \leq \exp\left\{\alpha_s\left(\beta_{sj}+\frac{\Delta\beta_{s(j+1)}}{2}\right)\right\}\nonumber\\
&\Longleftrightarrow&\frac{A_{sj}}{G_{sj}}\leq 
\frac{A_{sj}}{2u_{s(j+1)}}\left(\Delta u_{s(j+1)}\sqrt{1+\frac{4w_s\alpha_s\tilde x^{(k)}_s u_{s(j+1)}}{\Delta u_{s(j+1)\mu^sA_{sj}}}}-u_{s(j+1)}-u_{sj}\right)
\leq A_{sj}G_{s(j+1)}\nonumber\\
&\Longleftrightarrow&\frac{\left(\frac{2u_{s(j+1)}}{G_{sj}}+u_{sj}+u_{s(j+1)}\right)^2}{\Delta u_{s(j+1)}^2}\leq 1+\frac{4w_s\alpha_s\tilde x^{(k)}_s u_{s(j+1)}}{\Delta u_{s(j+1)}\mu^sA_{sj}}\leq \frac{\left(2u_{s(j+1)}G_{s(j+1)}+u_{sj}+u_{s(j+1)}\right)^2}{\Delta u_{s(j+1)}^2}\nonumber\\
\label{eq:1_rho_inequality}
&\Longleftrightarrow&\qquad\frac{\tilde x^{(k)}_s}{\mu^s}\geq A_{sj}\frac{\left(\frac{2u_{s(j+1)}}{G_{sj}}+u_{sj}+u_{s(j+1)}\right)^2-\Delta^2 u_{s(j+1)}}{4w_s\alpha_su_{s(j+1)}\Delta u_{s(j+1)}}\\
\label{eq:2_rho_inequality}
&\textrm{and}&\qquad\frac{\tilde x^{(k)}_s}{\mu^s}\leq A_{sj} 
\frac{\left[2u_{s(j+1)}G_{s(j+1)}+u_{sj}+u_{s(j+1)}\right]^2-\Delta^2 u_{s(j+1)}}{4w_s\alpha_su_{s(j+1)}\Delta u_{s(j+1)}}
\end{eqnarray}
For equation (\ref{eq:1_rho_inequality}), simple algebraic manipulations yields
\begin{eqnarray}
\frac{\tilde x^{(k)}_s}{\mu^s}&\geq&A_{sj}\frac{\left(\frac{2u_{s(j+1)}}{G_{sj}}+u_{sj}+u_{s(j+1)}\right)^2-\Delta^2 u_{s(j+1)}}{4w_s\alpha_su_{s(j+1)}\Delta u_{s(j+1)}}\nonumber\\
&=&\frac{A_{sj}}{4w_s\alpha_su_{s(j+1)}\Delta u_{s(j+1)}}\left(\frac{2u_{s(j+1)}}{G_{sj}}+u_{sj}+u_{s(j+1)}+\Delta u_{s(j+1)}\right)\left(\frac{2u_{s(j+1)}}{G_{sj}}+u_{sj}+u_{s(j+1)}-\Delta u_{s(j+1)}\right)\nonumber\\
&=&\frac{A_{sj}}{4w_s\alpha_su_{s(j+1)}\Delta u_{s(j+1)}}\left(\frac{2u_{s(j+1)}}{G_{sj}}+2u_{s(j+1)}\right)\left(\frac{2u_{s(j+1)}}{G_{sj}}+2u_{sj}\right)\nonumber\\
&=&\frac{A_{sj}}{w_s\alpha_s\Delta u_{s(j+1)}G_{sj}^2}\left(G_{sj}+1\right)\left(u_{s(j+1)}+G_{sj}u_{sj}\right)\nonumber
\end{eqnarray}
Similarly, for (\ref{eq:2_rho_inequality}) we get
\begin{eqnarray}
\frac{\tilde x^{(k)}_s}{\mu^s}&\leq&A_{sj} 
\frac{\left(2u_{s(j+1)}G_{s(j+1)}+u_{sj}+u_{s(j+1)}\right)^2-\Delta^2 u_{s(j+1)}}{4w_s\alpha_su_{s(j+1)}\Delta u_{s(j+1)}}\nonumber\\
&=&\frac{A_{sj}}{4w_s\alpha_su_{s(j+1)}\Delta u_{s(j+1)}} 
\left(\frac{}{}2u_{s(j+1)}G_{s(j+1)}+u_{sj}+u_{s(j+1)}+\Delta u_{s(j+1)}\right)\left(\frac{}{}2u_{s(j+1)}G_{s(j+1)}+u_{sj}+u_{s(j+1)}-\Delta u_{s(j+1)}\right)
\nonumber\\
&=&\frac{A_{sj}}{{4w_s\alpha_su_{s(j+1)}\Delta u_{s(j+1)}}} 
\left(2u_{s(j+1)}G_{s(j+1)}+2u_{s(j+1)}\right)\left(2u_{s(j+1)}G_{s(j+1)}+2u_{sj}\right)\nonumber\\
&=&\frac{A_{sj}}{w_s\alpha_s\Delta u_{s(j+1)}} 
\left(G_{s(j+1)}+1\right)\left(u_{s(j+1)}G_{s(j+1)}+u_{sj}\right)\nonumber
\end{eqnarray}
And finally, we obtain
\begin{eqnarray}
\frac{w_s\alpha_s\Delta u_{s(j+1)}}{A_{sj} 
\left(G_{s(j+1)}+1\right)\left(u_{s(j+1)}G_{s(j+1)}+u_{sj}\right)}\leq \frac{\mu^s}{\tilde x^{(k)}_s} \leq
\frac{w_s\alpha_s\Delta u_{s(j+1)}G_{sj}^2}{A_{sj}\left(G_{sj}+1\right)\left(u_{s(j+1)}+G_{sj}u_{sj}\right)}
\end{eqnarray}
which after substituting $y$ and $j$ respectively by $\tilde x_s^{(k+1)}$ and $i_s^{(k+1)}$, completes the proof.
\end{figure*}


\begin{thebibliography}{1}
\bibitem{schaar2005cross}
M. van Der Schaar and S. Shankar, ``Cross-Layer Wireless Multimedia Transmission: Challenges, Principles, and New Paradigms,'' IEEE Wireless Communications Magazine, vol. 12, no. 4, pp. 50–-58, 2005.

\bibitem{zhang2004end}
Q. Zhang, W. Zhu, and Y.-Q. Zhang, ``End-to-end QoS for Video Delivery over Wireless Internet,''
\emph{Proceedings of the IEEE}, vol. 93, no. 1, pp. 123--134, 2004.

\bibitem{ChangVideoAdaptation}
S. F. Chang and A. Vetro, ``Video adaptation: concepts, technologies, and open issues,'' \emph{Proceedings of the IEEE}, pp. 148-158, 2005.

\bibitem{OhmScalable}
J. Ohm, ``Advances in Scalable Video Coding,'' \emph{Proceedings of the IEEE}, vol. 93, no. 1, pp. 42-56, 2005.

\bibitem{SchwarzScalable}
H. Schwarz, D. Marpe, and T. Wiegand, ``Overview of the scalable video coding extension of H.264/AVC,'' \emph{IEEE Transactions on Circuits and Systems for Video Technology}, vol. 17, no. 9, pp. 1103-1120, 2007.

\bibitem{Dai:2003p3625}
M. Dai and D. Loguinov, ``Analysis of Rate-Distortion Functions and Congestion Control in Scalable Internet Video Streaming,'' in \emph{ACM NOSSDAV}, 2003, pp. 60--69.

\bibitem{XZhu}
X. Zhu and B. Girod, ``Distributed media-aware rate allocation for wireless video streaming,'' in \emph{Picture Coding Symposium (PCS09)}, Chicago, IL, USA, May 2009.

\bibitem{ZhuLiContentAware}
Y. Li, Z. Li, M. Chiang, and A. R. Calderbank, ``Content-aware distortion fair video streaming in networks,'' \emph{IEEE Transactions on Multimedia}, vol. 11, no. 6, pp. 1182--1193, 2009.

\bibitem{nejati2010distortion}
N. Nejati, H. Yousefi'zadeh, and H. Jafarkhani, ``Distortion Optimal Transmission of Multi-Layered FGS Video over Wireless Channels,'' \emph{IEEE Journal on Selected Areas in Communications}, vol. 28, no. 3, pp. 510--519, 2010.

\bibitem{Girod}
X. Zhu and B. Girod, ``Distributed Media-Aware Rate Allocation for Wireless Video Streaming,'' \textit{IEEE Transactions on Circuits and Systems for Video Technology}, vol. 20, no. 11, pp 1462--1474, 2010.

\bibitem{katsaggelos}
C. E. Luna, L. P. Kondi, and A. K. Katsaggelos, ``Maximizing User Utility in Video Streaming Applications,'' \textit{IEEE Transactions on Circuits and Systems for Video Technology}, vol. 13, no. 2, pp. 141--148, 2003.

\bibitem{num_satellite}
D. Pradas, M.A. Vazquez-Castro, ``NUM-Based Fair Rate-Delay Balancing for Layered Video Multicasting over Adaptive Satellite Networks,'' \emph{IEEE Journal on Selected Areas in Communications}, vol. 29, no. 5, pp. 969-978, 2011.

\bibitem{TVT}
J. Zou, H. Xiong, C. Li, R. Zhang, and Z. He ``Lifetime and Distortion Optimization With Joint Source/Channel Rate Adaptation and Network Coding-Based Error Control in Wireless Video Sensor Networks,'' \emph{IEEE Trans. on Vehicular Technology}, vol. 60, no. 3, pp. 1182-1194, 2011.

\bibitem{Kang:2007p3384}
S. R. Kang and D. Loguinov, ``Modeling Best-Effort and FEC Streaming of Scalable Video in Lossy Network Channels,'' \emph{IEEE/ACM Transactions on Networking}, vol. 15, no. 1, pp. 187-200, 2007.

\bibitem{Yan:2006p3580}
J. Yan, K. Katrinis, M. May, and B. Plattner, ``Media-and TCP-Friendly Congestion Control for Scalable Video Streams,'' \emph{IEEE Transactions on Multimedia}, vol. 8, no. 2, pp. 196-206, 2006.

\bibitem {Kelly}
F. P. Kelly, A. Maulloo, and D. K. H. Tan, ``Rate Control for Communication Networks: Shadow Prices, Proportional Fairness and Stability,'' \emph{Journal of Operations Research Society}, vol. 49, pp. 237-252, 1998.

\bibitem{Low}
S. H. Low and D. E. Lapsley, ``Optimization Flow Control, I: Basic Algorithm and Convergence,'' \emph{IEEE/ACM Transactions on Networking}, vol. 7, no. 6, pp. 861-874, 1999.

\bibitem{LayeringChiang}
M. Chiang, S. H. Low, A. R. Calderbank, and J. C. Doyle, ``Layering as Optimization Decomposition: A Mathematical Theory of Network Architectures,'' \emph{Proceedings of the IEEE}, vol. 95, no. 1, pp. 255-312, 2007.


\bibitem{Ya-Qin_Zhang}
D. Wu, Y. T. Hou, and Y.-Q. Zhang, ``Scalable Video Coding and Transport over Broadband Wireless Networks,'' \emph{Proceedings of the IEEE}, vol. 89, no. 1, pp. 6--20, 2002.

\bibitem{Shenker}
S. Shenker, ``Fundamental design issues for the future Internet,'' \emph{IEEE Journal on Selected Areas in Communications}, vol. 13, no. 7, pp. 1176-1188, 1995.

\bibitem{Shroff}
J. W. Lee, R. R. Mazumdar, and N. B. Shroff, ``Nonconvex optimization and rate control for multi-class services in the Internet,'' \emph{IEEE/ACM Transactions on Networking}, vol. 13, no. 4, pp. 827-840, 2005.

\bibitem{ChiangInelastic}
P. Hande, S. Zhang, and M. Chiang, ``Distributed rate allocation for inelastic flows,'' \emph{IEEE/ACM Transactions on Networking}, vol. 15, no. 6, pp. 1240-1253, 2007.

\bibitem{Talebi_ICC12}
A. Sehati, M. S. Talebi, A. Khonsari, ``NUM-Based Rate Allocation for Streaming Traffic via Sequential Convex Programming," in \emph{International Conference on Communications (ICC12)}, Ottawa, Canada, June 2012.

\bibitem{InelasticWSN}
J. Jin, A. Sridharan, B. Krishnamachari, and M. Palaniswami, ``Handling Inelastic Traffic in Wireless Sensor Networks,'' \emph{IEEE Journal on Selected Areas in Communications}, vol. 28, no. 7, pp. 1105-1115, 2010.

\bibitem{SLOW2}
M. H. Wang, M. Palaniswami, and S. H. Low, ``Application-Oriented Flow Control: Fundamentals, Algorithms and Fairness'', \emph{IEEE/ACM Transactions on Networking}, vol. 14, no. 6, pp. 1282-1291, 2006.


\bibitem{Talebi-Elsevier}
M. S. Talebi, A. Khonsari, and M. H. Hajiesmaili, ``Utility-Proportional Bandwidth Sharing for Multimedia Transmission Supporting Scalable Video Coding,'' \emph{Computer Communications}, vol. 33, no. 13, pp. 1543--1556, 2010.

\bibitem{Abbas}
G. Abbas, A. K. Nagar, H. Tawfik, and J. Y. Goulermas, ``Quality of service issues and nonconvex network utility maximization for inelastic services in the Internet,'' in \emph{IEEE/ACM MASCOTS}, London, UK, September 2009, pp. 537--547.

\bibitem{Tang}
M. Tang, C. Long, and X. Guan, ``Nonconvex maximization for communication systems based on particle swarm optimization'', \emph{Computer Communications}, vol. 33, no. 7, pp. 841--847, 2010.

\bibitem{Zegura}
Z. Cao and E. W. Zegura, ``Utility max-min: an application-oriented bandwidth allocation scheme,'' in \emph{IEEE INFOCOM}, 1999, pp. 793--801.

\bibitem{Talebi-GC}
M. S. Talebi, A. Khonsari, M. H. Hajiesmaili, and S. Jafarpour,
``A suboptimal network utility maximization approach for scalable multimedia applications,'' in \emph{IEEE Globecom'09}, 2009, pp. 5578--5583.


%


\bibitem{global_opt}
R. Horst and H. Tuy, \emph{Global optimization: deterministic approaches,} Springer-Verlag, 1993.

\bibitem{SCP_EE364}
S. Boyd, ``Sequential Convex Programming," EE364b Lecture Notes,
Stanford University.

\bibitem{SCP_TranDiehl}
Q. Tran Dinh and M. Diehl, ``Local convergence of sequential
convex programming for nonlinear optimization,'' \emph{In M. Diehl, F. Glineur, E. Jarlebring, and W. Michiels (eds.), Recent Advances in Optimization and its Applications in Engineering,} pp. 93--102, Springer-Verlag, Berlin, Heidelberg, 2010.

\bibitem{Bert_NLP}
D. P. Bertsekas, \emph{Nonlinear Programming,} Belmont, MA: Athena Scientific, 1999.

\bibitem{pso}
J. Kennedy and R. C. Eberhart, ``Particle swarm optimization,'' in \emph{IEEE International
Conference on Neutral Networks}, Perth, Australia, 1995.

\bibitem{QoE2}
W. Wu, A. Arefin, R. Rivas, K. Nahrstedt, R. Sheppard, and Z. Yang. ``Quality of experience in distributed interactive multimedia environments: Toward a theoretical framework,'' \emph{In Proc. ACM Multimedia}, 2009.


\bibitem{QoE}
ITU-T SG12, ``Definition of Quality of Experience,'' COM12 Ð LS 62 Ð E, TD 109rev2 (PLEN/12), Geneva, Switzerland, 16-25 Jan 2007.

\bibitem {Boyd}
S. Boyd and L. Vandenberghe, \emph{Convex Optimization,} Cambridge University Press, 2004.

\bibitem{Evans_TSP}
J. C. F. Li, S. Dey, and J. S. Evans, ``Maximal Lifetime Power and Rate Allocation for Wireless Sensor Systems With Data Distortion Constraints," \emph{IEEE Transactions on Signal Processing}, vol. 56, no. 5, May 2008, pp. 2076--2090.


\bibitem{Balancing_chiang}
M. Chiang, ``Balancing Transport and Physical Layers in Wireless Multihop Networks: Jointly Optimal Congestion Control and Power Control,'' \emph{IEEE Journal on Selected Areas in Communications}, vol. 23, no. 1, pp. 104-116, 2005.

\bibitem{SVCTest}
M. Mathew and H. Schwarz, ``Testing Conditions for SVC Coding Efficiency and JSVM Performance Evaluation,'' Joint Video Team of ISO/IEC MPEG and ITU-T VCEG, Doc. JVT- Q205, Oct. 2005.
	
\bibitem{SVC-sequence}
SVC test sequences. [Online]. Available: ftp.tnt.uni-hannover.de/pub/svc/testsequences/.





\bibitem{Wenger}
S. Wenger, Y. K. Wang, T. Shierl, ``Transport and Signaling of SVC in IP Networks,'' \textit{IEEE Trans. on Circuits and Systems for Video Tech.}, vol. 17, no. 9, Sep. 2007.



\end{thebibliography}
\end{document}